\documentclass[aps,pra,superscriptaddress,showkeys,floatfix,twocolumn,nobibnotes,nofootinbib]{revtex4}
\usepackage{hyperref}
\hypersetup{colorlinks,allcolors=black}

\usepackage{amsmath,amssymb}    
\usepackage{amsfonts} 
\usepackage{graphicx}   
\usepackage{verbatim}   
\usepackage{xcolor}     
\usepackage{subfigure}  

\usepackage[normalem]{ulem} 
\usepackage[super]{nth} 

\usepackage{pbox}

\newcommand{\eff}{\text{eff}}
\newcommand{\TE}{\text{TE}}
\newcommand{\TM}{\text{TM}}

\begin{document}
\title{Single scattering and effective medium  description in multilayer cylindrical metamaterials: Application to graphene  and  metasurface coated cylinders} 
\author{Charalampos P. Mavidis}
\email{mavidis@iesl.forth.gr}
\affiliation{Department of Materials Science and Technology, University of Crete, Heraklion, Crete, Greece}
\affiliation{Institute of Electronic Structure and Laser, Foundation for Research and Technology Hellas, N. Plastira 100, 70013 Heraklion, Crete, Greece}
\author{Anna C. Tasolamprou}
\email{atasolam@iesl.forth.gr}
\affiliation{Institute of Electronic Structure and Laser, Foundation for Research and Technology Hellas, N. Plastira 100, 70013 Heraklion, Crete, Greece}
\affiliation{Section of Electronic Physics and Systems, Department of Materials Science and Technology, University of Crete, Heraklion, Crete, Greece}

\author{Eleftherios N. Economou}
\affiliation{Institute of Electronic Structure and Laser, Foundation for Research and Technology Hellas, N. Plastira 100, 70013 Heraklion, Crete, Greece}
\affiliation{Department of Physics, University of Crete, Heraklion, Greece}

\author{Costas M. Soukoulis}
\affiliation{Institute of Electronic Structure and Laser, Foundation for Research and Technology Hellas, N. Plastira 100, 70013 Heraklion, Crete, Greece}
\affiliation{Ames Laboratory and Department of Physics and Astronomy, Iowa State University, Ames, Iowa 50011, USA}
\author{Maria Kafesaki}
\affiliation{Department of Materials Science and Technology, University of Crete, Heraklion, Crete, Greece}
\affiliation{Institute of Electronic Structure and Laser, Foundation for Research and Technology Hellas, N. Plastira 100, 70013 Heraklion, Crete, Greece}

\begin{abstract}
Coated and multicoated cylinder systems constitute an appealing metamaterial category, as they allow a very rich and highly tunable response, resulting from the interplay of the many different geometrical and material parameters involved. Here we derive and propose an effective medium approach for the detailed description and analysis of the electromagnetic wave propagation in such systems. In particular, we investigate infinitely-long multilayered cylinders with additional electric and magnetic surface conductivities at each interface. Our effective medium approach is based on the well known in the solid state physics community Coherent Potential Approximation  (CPA) method, combined with a transfer matrix-based formulation for cylindrical waves.
Employing this effective medium scheme, we investigate two realistic systems, one comprising of cylindrical tubes made  of uniform tunable graphene sheets and one of  cylinders/tubes formed  of metasurfaces exhibiting both electric and magnetic sheet conductivities. Both systems show a rich palette of engineerable electromagnetic features, including tunable hyperbolic response, double negative response and epsilon-near-zero and mu-near-zero response regions.

\end{abstract}

\maketitle

\section{Introduction}
\label{sec:intro}

\begin{figure*}[tb]
  \begin{center}
  \includegraphics{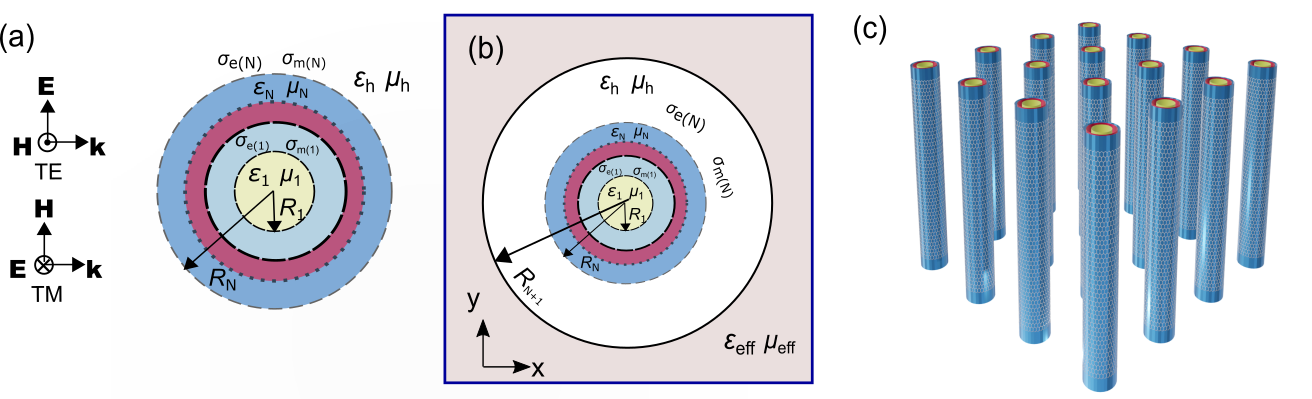}
    \caption{\label{fig:cylinder}
(a) Top view schematic of the more general meta-atom geometry investigated in this work: A cylinder of $N$ co-centric bulk layers, separated by metasurfaces. Each bulk layer is characterized by relative electric permittivity $\varepsilon_\ell$ and magnetic permeability $\mu_\ell$. The interface between the $\ell$-th and the $(\ell+1)$-th layer is coated with a metasurface with  electric conductivity $\sigma_{e(\ell)}$ and magnetic conductivity $\sigma_{m(\ell)}$. The polarization definition is also shown, where we assume normal incidence. (b) Schematic of the setup for the effective medium derivation. The original cylinder (of panel (a)) is coated with a cylindrical layer of the host material (white color), with radius $R_{N+1}=R_N/\sqrt{f}$, where $f$ is the filling ratio of the cylinders in the metamaterial under investigation, and is embedded  in the homogeneous effective medium under determination (with relative permittivity $\varepsilon_\eff$ and permeability $\mu_\eff$). (c) Three-dimensional view of the metamaterial under investigation: An array of multilayered cylinders.}
\end{center}
\end{figure*}

Electromagnetic metamaterials are artificial, structured  materials comprising of subwavelength resonant  building blocks, the meta-atoms. Due to their versatile nature, metamaterials  offer the possibility of novel and unconventional electromagnetic wave control, and thus advancements in a large variety of wave-control-related applications, including imaging, sensing, communications, etc.~\cite{Zhang2008NatMat,Soukoulis2011NatPhot,Lee2017Sensors,Liaskos2018IEEE}. Metamaterials' exceptional electromagnetic properties  stem to a larger degree from the architecture of the meta-atoms; through this architecture, the distribution of the local currents excited by an impinging electromagnetic wave is engineered, providing the desired response.   Particularly known forms of meta-atoms are properly aligned metallic short wires, behaving as macroscopic resonant electric dipoles and  producing a resonant electric response (resonant permittivity), and metallic spilt ring resonators, leading to resonant circulating currents and the emergence of a resonant magnetic response~\cite{Soukoulis2008JPCM}. Another approach to create resonant electric and/or magnetic response is by exploiting the Mie-based resonances in high index dielectric (or semiconducting) meta-atoms; this approach is typically proposed for applications in high (IR and optical) frequencies~\cite{Peng2007PRL,Kuznetsov2016Science}, where metals experience detrimentally high losses. Through tunable resonant magnetic and/or electric response one can engineer a plethora of different and peculiar metamaterial properties, such as negative, near-zero permittivity and/or permeability, negative refractive index,  peculiar anisotropy, asymmetric effects  and many more~\cite{Ferrari2015PCE,Liu2016,Basharin2013PRB,Katsantonis2020PRB} (Note that because of the subwavelength meta-atom size metamaterials provide homogeneous-medium-like (effective) properties and response).

Besides bulk (three-dimensional) metamaterials, many additional exciting    functionalities stem from the electromagnetic wave interaction with thin meta-atom layers, known as metasurfaces, which attract a constantly growing research attention. Metasurfaces, by allowing modulation of the meta-atoms along them, allow the engineering of both phase and amplitude of the electromagnetic fields impinging on them, acquiring thus the ability to replace bulk, heavy and difficult to use conventional optical elements (mirrors, lenses, etc.). Due to their ultrathin nature and the subwavelentg meta-atom size, metasurfaces can be conveniently described as effective electromagnetic sheets ~\cite{Glybovski2016PhysRep}, through appropriate sheet conductivities. Metasurfaces comprising of a thin layer sustaining orthogonal electric and magnetic dipoles have been utilized for applications as reflect-arrays, transmit-arrays, holographic surfaces and others  ~\cite{ChenKimWongEleftheriades}. Moreover, metasurfaces' fine electromagnetic features have been shown to enable enhanced detection and sensing, thin film polarizers, shielding, beam shaping  and other useful functionalities~\cite{Beruete2020AOM,Tasolamprou2014OptExpress,Skoulas2021,Tasolamprou2022,Zhu2021NatureComm,Perrakis2021SciRep}.

As mentioned, metamaterials and metasurfaces can be made of metallic, dielectric or semiconducting components. They can be also made of a combination of dielectric, semiconducting and metallic parts in a properly designed meta-atom architecture and cluster arrangement. A scheme that has gained significant popularity is structures composed of  coated (or even multicoated) cylinders or spheres. Such structures are characterized by a relatively straightforward design and have been proposed for a variety of applications due to the increased degree of design freedom related to the thicknesses and constituent  materials in each layer. With proper selection of geometry and materials, coated cylinders or spheres can lead to  overlapping of different resonances, which is crucial in metasurfaces since it can offer full  transmission or reflection and $2\pi$ phase modulation (allowing in principle arbitrary wavefront control),  resonances with engineered quality factors, etc. Applications of such structures include superscattering~\cite{Ruan2010PRL,Lepeshov2019ACSphot,Wu2019PRB,Raad2019JPD,raad2019multi,Abrashuly2019} and electromagnetic cloaking~\cite{Alu2005PRE,Bernety2015JPCM,Zhang2016OE,Naserpour2016SciRep,Labate2017PRA,Batool2020Photonics,Raad2019JPD,Shcherbinin2020PhysRevA,Zheng2021OL},  lenses and many others~\cite{Lock2008JOSA}. 
  Moreover, metamaterials made of cylindrical meta-atoms, which are the system of  interest in the present work, are inherently anisotropic, allowing the possibility of hyperbolic dispersion relation and anisotropic negative or near-zero refractive index~\cite{Kumar2020OE}. Such structures  can be experimentally realized following the progress of micro and nanotechnology; for example  emerging technologies focused on the implementation of carbon nanotubes have given already metamaterial and photonic crystal orientated developments~\cite{Kaplas2017,Kuzhir2021484,Shuba2020,Sedelnikova2021}.
 Even more interesting electromagnetic features can occur in cylindrical  meta-atoms coated with tunable sheets bearing individual electric and/or magnetic  resonances to be combined with the response of the coated atom. Such coatings may involve, for example, a 2D material, like uniform graphene or structured (patterned) graphene, or an electromagnetically thin sheet of cut wires or split ring resonators~\cite{Alu2005PRE,Chen2013,Shcherbinin2020PhysRevA,Christensen2015,Tkachova20191021}, e.g. in a flexible  metasurface implementation~\cite{Qian20181231,Tasolamprou2020f}. It should be mentioned here that graphene in particular, either in a patterned metasurface form or as a uniform sheet, is very appealing as a coating material due to its intrinsic ultrathin nature and the exceptionally tunable electromagnetic properties, especially in the THz wavelengths where its EM behavior is dominated by a Drude-like response~\cite{Yao20161518,Tasolamprou2019ACSPhot,Ahmadivand20198091,Koulouklidis20223075}. 

It is clear that an analytical  assessment of the electromagnetic response of coated and multicoated cylinder-based metamaterial structures is important, as it gives the possibility for the in-depth understanding of the physical mechanisms that lead to the resonant structure response, and, subsequently, for engineering of this response through structure engineering and optimization, targeting advanced electromagnetic functionalities and applications. Assemblies of resonant cylinders can be treated as an effective homogeneous material in the limit of small characteristic lengths (radius, unit cell size) compared to the wavelength of interest. 
Homogenization approaches applied in systems of coated spheres and cylinders have shown that coatings can provide many interesting effects, as for example an increased bandwidth of negative permittivity and permeability in comparison with their non-coated counterparts~\cite{Yan2012PER}. 
However, to our knowledge, an analytic homogenization approach that can incorporate an arbitrary number (larger than one) of coatings for each cylinder has not been reported in the literature yet. Additionally, although the scattering properties of cylinders and spheres coated with graphene metasurfaces have been quite extensively studied, coatings/sheets showing arbitrary resonant electric and/or resonant magnetic response (representing more complex metasurface-coatings and allowing delicate interplay of electric and magnetic dipoles, resulting to additional advanced functionalities)~\cite{She2007OE,Alu2009PRB,Danaeifar2017JOSAB,Shokati2017AOP,Zarghani2019OE,Tassin2012PhysB,Tsilipakos2018ACSphot,Droulias2020PRB} are much less explored.

The aim of this work is to develop a framework/formalism to analyze in detail the resonant behaviour and wave
propagation in systems of multilayer cylinders coated also with conducting sheets (metasurfaces) of both electric and magnetic response, as well as to apply this framework to cases of high foreseen theoretical or practical interest.  Towards this direction, we derive an homogeneous effective medium approach for systems of infinitely-long multilayer cylinders, with an arbitrary number of layers (coatings; of metallic, high-index dielectric or even resonant materials) and with the incorporation  of both electric and magnetic sheet conductivities at each interface (i.e. between coatings).  Our homogenization approach is based on the well known in the solid state physics community Coherent Potential Approximation, CPA~\cite{Kafesaki19977,Kafesaki1998383,Wu2006PRB,PSheng2006book,Mavidis2020PRBpol}, suitable also quite beyond the long-wavelength limit. To calculate the single cylinder scattering amplitudes for the multilayer cylinders, which is an essential step in the CPA application, we develop a Transfer Matrix Method (TMM) for cylindrical waves, connecting the wave amplitudes at the different layers. We apply the developed formalism in two different systems/metamaterials: (i) of cylinders (nanotubes) of  uniform graphene sheets, with tunable response, and (ii) of cylinders formed of  metasurfaces with arbitrary electric and magnetic resonances in the metasurface conductivity. In both systems we study the single meta-atom (multi-coated cylinder) scattering and the effective medium response, which unveils the existence of rich electromagnetic features, i.e., controllable  hyperbolic response of both type I and type II, double negative response, and epsilon-near-zero and mu-near-zero response. 
The paper is organized as follows: In Sec.~\ref{sec:metasurface-methods} we present our method, i.e., starting from the single meta-atom scattering we derive the relations for the effective  electric permittivity and magnetic permeability tensor components for the corresponding metamaterial.
In Sections~\ref{subsc:results-singlesc} and \ref{subsc:results-emt} we apply the method in systems of single- and double-layer/wall cylindrical nanotubes made of (a) tunable graphene sheets and (b) metasurfaces with both electric and magnetic   surface conductivity. The first case approximates systems of single and double-wall carbon nanotubes, which are systems of high technological interest. The second system can approximate, among others, metamaterials of cylinders coated with a structured 2D material, e.g. graphene, transition metal dichalcogenide monolayers, etc.    In both cases we demonstrate the  engineerable effective electric permittivity and magnetic permeability response, leading to the emergence of various interesting optical phases and possibilities.   Finally we present the conclusions on our work.



\section{Methods}
\label{sec:metasurface-methods}
We begin our analysis from the methods derived and  employed in this work for the system shown in Fig.~\ref{fig:cylinder}.
In the first part we present the THz electric and magnetic sheet conductivities of the metasurfaces considered in the application cases discussed  here. In the  second part we present first the derivation of a Transfer Matrix Method, which allows us to calculate the scattering properties of a cylinder composed of $N$ co-centered layers of different materials, with metasurfaces at the interfaces of these layers. Next, we derive the CPA-based effective medium model for two-dimensional arrays of such multilayered cylinders, based on the single scattering calculations.
\subsection{2D Conductivities}
\label{subsc:2D-conductivities}

In this subsection we present the electromagnetic properties of the 2D sheets/coatings of the examples considered in this work, i.e. the uniform graphene sheet and the metasurface exhibiting both electric and magnetic resonance. 

\begin{figure}[hhh]
  \begin{center}
   \includegraphics[width=86mm]{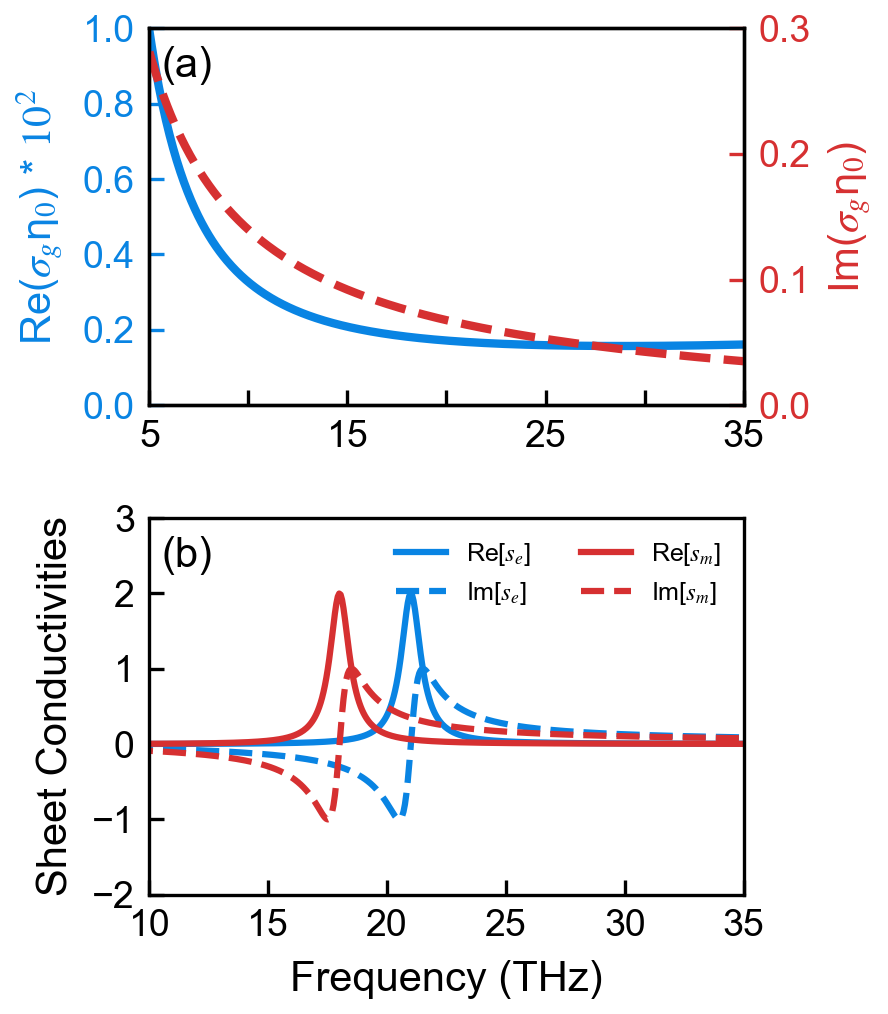}
    \caption{\label{fig:graphene-cond} (a) Real part (times 100) (left axis, blue line) and imaginary part (right axis, red line) of the sheet conductivity of a uniform graphene sheet  modelled by  Eq.~\eqref{eq:graphene_rpa2}, assuming Fermi level $E_F=0.2$~eV and relaxation time $\tau=1$~ps. (b) Normalized real (solid lines) and imaginary  (dashed lines) parts of electric ($s_e$) and magnetic ($s_m$) sheet conductivities calculated using Eqs.~\eqref{eq:sigma_e}-\eqref{eq:sigma_m}, assuming parameters $\omega_e/2\pi=21$~THz, $\Gamma_e/2\pi=\Gamma_m/2\pi=2$~THz,  $\kappa_e/2\pi=\kappa_m/2\pi=1$~THz and $\omega_m/2\pi=18$~THz; for the definition of those parameters see main text. $\eta_0$ is the free-space impedance.
}
  \end{center}
\end{figure}

For the graphene case the conductivity, $\sigma_g$, as a function of the Fermi energy, $E_F$, and the temperature, $T$,  was obtained by Kubo formula,  derived in the context of Rapid Phase Approximation (RPA)~\cite{RPA}; it reads as
\begin{equation}
\label{eq:graphene_rpa2}
\sigma_g(\omega) = \sigma_{\textrm{intra}}+\sigma_{\textrm{inter}}
\end{equation}
where the intraband contribution is
\begin{equation}
\sigma_{\textrm{intra}}(\omega) = \frac{2e^2 k_B T}{\pi \hbar^2} \frac{i}{\omega+i\tau^{-1}} \ln\left[2\cosh\left(\frac{E_F}{2 k_B T}\right)\right]
\label{eq:graphene-intra}
\end{equation}
and the interband contribution is
\begin{align}
\nonumber
\sigma_{\textrm{inter}}(\omega) =   \frac{e^2}{4\hbar} (\frac{1}{2}+\frac{1}{\pi}\arctan\left(\frac{\hbar\omega- 2 E_F}{2k_B T}\right) \\-\frac{i}{2\pi}\ln\left[\frac{(\hbar\omega+ 2 E_F)^2}{(\hbar\omega- 2 E_F)^2+(2k_B T)^2} \right]). 
\end{align}
Here $\omega$ is the angular frequency, $\hbar=1.055\times 10^{-34}$ Js the reduced Planck constant, $k_B=1.38 \times 10^{-23} \textrm{J}\cdot \textrm{K}^{-1}$ the Boltzmann constant, $e=1.602 \times 10^{-19}$ C the electron charge and $\tau$ the electrons' relaxation time.
Unless otherwise stated, in this paper we use a Fermi level $E_F=0.2$~eV and a typical relaxation time $\tau=1$~ps. The real and imaginary part of the graphene conductivity for these values and for our frequency region of interest are shown in Fig.~\ref{fig:graphene-cond}(a).

Regarding the metasurface coatings employed, usually in the literature their response is approximated by a sheet material with effective electric and magnetic 2D conductivities  consisting of a summation of Lorentzian resonances~\cite{Tassin2012PhysB,Radi2014IEEE,Mohammadi2016PRX,Tsilipakos2018ACSphot,Droulias2020PRB}. 
To simplify our analysis we consider here a metasurface with isotropic surface conductivities and a single electric and magnetic resonance:
\begin{equation}
\label{eq:sigma_e}
   s_e = \sigma_e\eta_0 =  \frac{i\kappa_e\omega}{\omega^2-\omega_{e}^2+i\omega\Gamma_e}
\end{equation}

\begin{equation}
\label{eq:sigma_m}
    s_m = \frac{\sigma_m}{\eta_0} =   \frac{i\kappa_m\omega}{\omega^2-\omega_{m}^2+i\omega\Gamma_m}
\end{equation}
where $\eta_0$ is the free-space impedance, $\omega_{e/m}$ are the resonance frequencies,  $\kappa_{e/m}$ and $\Gamma_{e/m}$ are parameters of the lorentzians. 
For the purpose of the present analysis we have chosen the values of the parameters involved in the  conductivities as $f_e = \omega_e/2\pi=21$~THz, $\Gamma_e/2\pi=\Gamma_m/2\pi=1$~THz, $\kappa_e/2\pi=\kappa_m/2\pi=2$~THz and $f_m=\omega_m/2\pi=18$~THz. The  real and imaginary part of the corresponding electric and magnetic conductivities  are shown in Fig.~\ref{fig:graphene-cond}(b).
\subsection{Single Scattering}
\label{subsec:metasurface-methods-singlescatt}
Having defined the sheet conductivities of the graphene and metasurface coatings  we move to the investigation of a single cylinder system. We consider an infinitely-long cylinder consisting of $N$ co-centered layers. The system is shown in Fig.~\ref{fig:cylinder}(a).
The $\ell$-th layer is characterized by its thickness $\Delta_\ell=R_\ell-R_{\ell-1}$ ($R_0=0$), where $R_\ell$ is the distance from the center to the perimeter of the $\ell$-th layer; its relative electric permittivity is $\varepsilon_\ell$, the relative magnetic permeability is $\mu_\ell$, the electric surface conductivity is $\sigma_{e(\ell)}$ and the  magnetic surface conductivity is $\sigma_{m(\ell)}$. The cylinder is embedded in a host material with relative electric permittivity $\varepsilon_h$ and magnetic permeability $\mu_h$. We consider wave propagation perpendicular to the cylinder axis. 
Since the cylinder is infinitely-long and there is no propagation component parallel to its axis, the problem  is
two dimensional and can be decoupled
into two separate polarizations, the transverse electric (TE), with the electric field normal to the cylinder
axis, and the transverse magnetic (TM) polarization, with the
magnetic field normal to the cylinder axis.
In each layer the fields can be expanded on the basis of cylindrical vector harmonics. 
In the $\ell$-th layer the field $\mathbf{F}=\lbrace\mathbf{E}, \mathbf{H}\rbrace$ (electic or magnetic) parallel to the cylinder axis ($z$ direction) will be 
\begin{equation}
    \mathbf{F}_\ell \sim \sum_{\nu}\left[c_{\ell\nu}\mathbf{N}_{e\nu k_\ell}^{(\textrm{outward})} + d_{\ell\nu}\mathbf{N}_{e\nu k_\ell}^{(\textrm{inward})}\right],
\end{equation}
 with  $\mathbf{N}_{e\nu k_\ell}^{(\textrm{outward})}\sim H_\nu(k_\ell r)$ and  $\mathbf{N}_{e\nu k_\ell}^{(\textrm{inward})}\sim J_\nu(k_\ell r)$ standing for the outgoing and ingoing cylindrical harmonics respectively~\cite{Stratton2015Book}.  The functions $J_\nu(\cdot)$ and $H_\nu(\cdot)$ are the Bessel and the first kind Hankel function of order $\nu$, and $k_\ell=\sqrt{\varepsilon_\ell \mu_\ell}\omega/c$.
The expansion coefficients $c_{\ell \nu}$ and $d_{\ell\nu}$ can be calculated by imposing the appropriate boundary conditions at the interface $r=R_{\ell}$~\cite{Kuester2003IEEE,Holloway2011IEEE,Dehmollaian2019IEEE}: 
\begin{align}
\label{eq:bc_vector}
\hat{\rho}\times\left[\mathbf{E}_{\ell+1} - \mathbf{E}_{\ell}\right] &= -\mathbf{j}_{m(\ell)} = -\sigma_{m(\ell)} \frac{\mathbf{H}_\ell + \mathbf{H}_{\ell+1}}{2} \\
\hat{\rho}\times\left[\mathbf{H}_{\ell+1} - \mathbf{H}_{\ell}\right] &= \mathbf{j}_{e(\ell)} = \sigma_{e(\ell)} \frac{\mathbf{E}_\ell + \mathbf{E}_{\ell+1}}{2}  \label{eq:bc_vector2}
\end{align}
where $\hat{\rho}$ is the unit vector along the radial direction. 
Here we have chosen a set of Bessel functions for our descriptions that is not linearly independent, i.e. $J_\nu(\cdot)$ and $H_\nu(\cdot)$ instead of $J_\nu(\cdot)$ and $Y_\nu(\cdot)$ that are commonly used in the literature, because it is more convenient for the effective medium description in Section~\ref{subsc:EMT}.

By applying the boundary conditions, Eq.~\eqref{eq:bc_vector} and Eq.~\eqref{eq:bc_vector2}, at each of the interfaces of the $N$ layers of the cylinder, we construct a matrix equation which connects the fields in the innermost layer with the fields outside the cylinder (incident plus scattered field), for each cylindrical wave/harmonic ($\nu$) excited. This Transfer Matrix equation reads: 
\begin{equation}
\label{eq:cyl-tmm}
\mathbb{M}_{(N),\nu}^\text{P}
\begin{pmatrix}
b_\nu \\[6pt]
0
\end{pmatrix} = 
\begin{pmatrix}
1 \\[6pt]
a_\nu
\end{pmatrix}
\end{equation}
where $\mathbb{M}_{(N),\nu}^\text{P}$ is the total transfer matrix for polarization $\mathsf{P}=\lbrace\TE,\TM\rbrace$. 
With $a_\nu \equiv c_{(N+1),\nu}$ we denote the scattering coefficient of the scattered wave (for coefficient $1 \equiv d_{(N+1),\nu}$  of the incident wave) and with $b_\nu=d_{1 \nu}$ the coefficient (inwards) for the core (innermost) layer, while $\nu$ stands for the excited mode (cylindrical harmonic).

The total transfer matrix $\mathbb{M}_{(N),\nu}^\text{P}$ is derived through the transfer matrices connecting the fields at neighboring layers.   
For polarization $\mathsf{P}=\lbrace\TE,\TM\rbrace$  the transfer matrix $\mathbb{T}_{\ell \nu}^\text{P}$ which transfers the fields from the $(\ell)$-th layer to the the fields in the $(\ell+1)$-th layer can be written as
\begin{equation}
\mathbb{T}_{\ell \nu}^\text{P}
\begin{pmatrix}
d_{\ell \nu} \\[6pt]
c_{\ell \nu}
\end{pmatrix} = 
\begin{pmatrix}
d_{(\ell+1),\nu} \\[6pt]
c_{(\ell+1),\nu}
\end{pmatrix}
\end{equation}

For TE polarization the matrix $\mathbb{T}_{\ell \nu}^\text{TE}$ has the form
\begin{equation}
    \mathbb{T}_{\ell \nu}^\TE = [\mathbb{D}_{(\ell+1) \nu}^\TE(R_\ell)]^{-1}\cdot  [\mathbb{X}_\ell^+]^{-1}\cdot\mathbb{X}_{\ell}^-\cdot \mathbb{D}_{\ell\nu}^\TE(R_{\ell}) 
\end{equation}
where
\begin{equation}
\mathbb{D}_{\ell \nu}^\text{TE}(R_\ell) =    \begin{pmatrix}
J'_\nu(k_\ell R_\ell)  & H'_\nu(k_\ell R_\ell)  \\[6pt]
\frac{1}{\eta_\ell} J_\nu(k_\ell R_\ell) & \frac{1}{\eta_\ell} H_\nu(k_\ell R_\ell)
\end{pmatrix}.   
\end{equation}
and the surface conductivity matrices ($ \mathbb{X}$) are
\begin{equation}
\mathbb{X}_{\ell}^\pm =    \begin{pmatrix}
1 & \pm i\sigma_{m(\ell)}/2\eta_0 \\ 
\mp i\sigma_{e(\ell)} \eta_0/2 & 1
\end{pmatrix}   
\end{equation}
where $\eta_\ell = \sqrt{\mu_\ell/\varepsilon_\ell}$ is the impedance of the $\ell$-th layer and $\eta_0=\sqrt{\mu_0/\varepsilon_0}$ is the vacuum impedance.

For TM polarization we get
\begin{equation}
    \mathbb{T}_{\ell \nu}^\TM = [\mathbb{D}_{(\ell+1) \nu}^\TM(R_\ell)]^{-1}\cdot  [\mathbb{X}_\ell^-]^{-1}\cdot\mathbb{X}_{\ell}^+\cdot \mathbb{D}_{\ell\nu}^\TM(R_{\ell}) 
\end{equation}
\begin{equation}
\mathbb{D}_{\ell \nu}^\text{TM}(R_\ell) =    \begin{pmatrix}
J_\nu(k_\ell R_\ell)  & H_\nu(k_\ell R_\ell)  \\[6pt]
\frac{1}{\eta_\ell} J'_\nu(k_\ell R_\ell) & \frac{1}{\eta_\ell} H'_\nu(k_\ell R_\ell)
\end{pmatrix} 
\end{equation}
The details of the calculations are presented in the Appendix A 1. 

The total transfer matrix reads as
\begin{equation}
   \mathbb{M}_{(N),\nu}^\mathtt{P} =  \prod_{\ell=N}^{1} \mathbb{T}_{\ell\nu}^\mathtt{P} 
   \label{eq:total-tranfer-matrix}
\end{equation}
From Eq.~\eqref{eq:cyl-tmm} we can calculate the coefficients $b_\nu$ (of the field in the core layer) and $a_\nu$ (scattered field coefficient) as
\begin{equation}
b_\nu = \frac{1}{\mathbb{M}_{(N),\nu}^{(11)}},
\end{equation}
\begin{equation}
 a_\nu = \mathbb{M}_{(N),\nu}^{(21)}b_\nu = \frac{\mathbb{M}_{(N),\nu}^{(21)}}{\mathbb{M}_{(N),\nu}^{(11)}}.
 \label{eq:sc-coeff}
\end{equation}
The scattering and extinction efficiencies of the whole cylinder can all be written in terms of $a_\nu$ as
\begin{equation}
    Q_{\textrm{ext}}^{\texttt{P}} = -\frac{2}{|k_h R_{N}|}\textrm{Re}\left[ a_0^{\texttt{P}} + 2\sum_{\nu=1}^{\infty} a_\nu^{\texttt{P}}  \right],
    \label{eq:Qext2}
\end{equation}
\begin{equation}
    Q_{\textrm{sc}}^{\texttt{P}} = \frac{2}{|k_h R_{N}|}\left[ |a_0^{\texttt{P}} |^2 + 2\sum_{\nu=1}^{\infty} |a_\nu^{\texttt{P}} |^2 \right].
    \label{eq:Qsc2}
\end{equation}

{\em Limiting expressions for a metasurface-covered cylinder:}
Having the above equations, one can derive limiting expressions for different systems of practical or theoretical interest. Here, we derive expressions for the resonance frequencies (poles) of the $\nu=1$ mode of a single-layer cylinder coated with a metasurface.
For a single cylindrical layer ($N=1$) with radius $R_1=R$ coated with a surface with conductivities $\sigma_e$ and $\sigma_m$ in a host material with electric permittivity $\varepsilon_h$ and magnetic permeability $\mu_h$, the scattering coefficients will be given based on Eq.~\eqref{eq:sc-coeff}.
In the limit $k_h R\ll 1$  and ignoring terms containing the product term $\sigma_e\sigma_m$, the poles of the  TE$_1$ mode can be found from the expression
\begin{equation}
\frac{1}{\eta_1}\frac{J_1(k_1R)}{J'_1(k_1R)} = \frac{\sigma_e\eta_0\eta_h-ik_hR}{i\eta_h+k_h R \sigma_m\eta_0^{-1}} 
\end{equation}
Further, if we take the quasistatic limit of $k_hR\ll 1$ and  $k_1 R\ll 1$, we find 
\begin{equation}
\label{eq:te1-mode-single}
    \varepsilon_1\frac{\omega}{c}R = \frac{ i\sigma_e\eta_0+\varepsilon_h\frac{\omega}{c}R}{\varepsilon_hi\sigma_m\eta_0^{-1}\frac{\omega}{c}R-1 }
\end{equation}
or
\begin{equation}
    \varepsilon_1 \varepsilon_h i\sigma_m\eta_0^{-1} \left(\frac{\omega}{c}R\right)^2- (\varepsilon_1+\varepsilon_h)\frac{\omega}{c}R  = i\sigma_e\eta_0 \label{eq:te1-lim1}
\end{equation}
For the sake of our analysis we ignore the damping term in the conductivities  (see Eqs.~\eqref{eq:sigma_e}-\eqref{eq:sigma_m}), i.e. we consider
\begin{equation}
    s_{e/m} = \frac{i\kappa_{e/m}\omega}{\omega^2-\omega^2_{e/m}} 
    \label{eq:lorentzian-cond}
\end{equation}
For $x=k_0R=\omega R/c \ll 1$ we can write the magnetic sheet conductivity as
$s_m = \sigma_m \eta_0^{-1} \simeq i\kappa_m (-(c/R)\cdot x/\omega^2_m-(c/R)^3\cdot  x^3/\omega_m^4) + \mathcal{O}(x^4)$,
and hence, can ignore the first term of Eq.~\eqref{eq:te1-lim1}. In this case, we can use Eqs.~\eqref{eq:te1-lim1} and \eqref{eq:lorentzian-cond} to get:
\begin{equation}
- (\varepsilon_1+\varepsilon_h)\frac{\omega}{c}R  = i\frac{i\kappa_{e}\omega}{\omega^2-\omega^2_{e}} 
\end{equation}
or
\begin{equation}
- (\varepsilon_1+\varepsilon_h)\frac{\omega}{c}R (\omega^2-\omega^2_{e})  = -\kappa_{e}\omega
\end{equation}
or
\begin{equation}
 (\varepsilon_1+\varepsilon_h)\frac{R}{c} (\omega^2-\omega^2_{e})  = \kappa_{e}.
\end{equation}
Finally we find that the  frequency of the TE$_1$ resonance of the structure is at  
\begin{equation}
    \omega^2_{\textrm{TE}_1} \simeq  \omega_e^2 + \frac{c\kappa_{e}}{(\varepsilon_1+\varepsilon_h)R}.
    \label{eq:te1-lim}
\end{equation}
An equivalent expression can be obtained for the TM$_1$ resonance: 
\begin{equation}
    \omega^2_{\textrm{TM}_1} \simeq  \omega_m^2 + \frac{c\kappa_{m}}{(\mu_1+\mu_h)R}.
    \label{eq:tm1-lim}
\end{equation}

For graphene, if we ignore the interband conductivity term in Eq.~\eqref{eq:graphene_rpa2}, the sheet conductivity takes the form $\sigma_g(\omega)\simeq i\kappa_g/\omega$, where $\kappa_g = \frac{2e^2 k_B T}{\pi \hbar^2} \ln\left[2\cosh\left(\frac{E_F}{2 k_B T}\right)\right]$; we then get a $1/\sqrt{R}$ dependence of the TE$_1$ mode resonance frequency.
\subsection{Effective Medium Theory}
\label{subsc:EMT}
In this section we derive the components of the effective  permittivity and permeability tensors for a uniaxial anisotropic system/metamaterial of infinitely-long parallel circular multicoated cylinders of $N$ layers each and surface electric and magnetic conductivities at each cylinder interface. 

We follow the same approach as the one of Refs.~\cite{Wu2006PRB,Mavidis2020PRBpol}, where (in ~\cite{Mavidis2020PRBpol}), we calculated the effective medium parameters for a cluster of  cylinders without coating and surface conductivities.
To derive the effective medium equations in the case of $N$-coated cylinders we consider a cylinder of
 $N$+1 layers embedded in the effective medium, as depicted in Fig.~\ref{fig:cylinder}(b), and we require the vanishing of the scattering amplitudes.  
The $(N+1)$-th layer of that cylinder is the host of the original system (with $\varepsilon_{N+1}=\varepsilon_h$ and $\mu_{N+1}=\mu_h$) with thickness $R_{N+1}-R_N$. The radius $R_{N+1}$ of the outer layer is determined by the filling ratio, $f$, of the cylinders in the original system, as $f=R_N^2/R_{N+1}^2$, i.e. the host-coated cylinder preserves the filling ratio of the original system/metamaterial. 
The host material (effective medium)  in the configuration of  Fig.~\ref{fig:cylinder}(b) has permittivity $\varepsilon_{\text{eff}} = \varepsilon^\perp_{\text{eff}}\left(\hat{x}\hat{x}+\hat{y}\hat{y} \right)+\varepsilon^\parallel_{\text{eff}}\hat{z}\hat{z}$ and permeability $\mu_{\text{eff}}= \mu^\perp_{\text{eff}}\left(\hat{x}\hat{x}+\hat{y}\hat{y} \right)+\mu^\parallel_{\text{eff}}\hat{z}\hat{z}$ (the symbols $\parallel, \perp$ are defined relative to the cylinders axes, i.e. for $\varepsilon^\parallel$ ($\mu^\parallel$) electric (magnetic) field is parallel to the axes of the cylinders).
In order to derive the expressions for the tensor components of the effective permittivity and permeability we require that the scattering coefficients of the scattered field in the effective medium, $a_\nu^{(\eff)}$,  for both TE and TM polarization vanish.
After algebraic manipulations (see details in Appendix~\ref{app2:EMT}), this requirement leads to expressions for the coefficients of the original cylinder, which read as 
\begin{equation}
\label{eq:generalized-cpa-index}
    a_\nu^\text{P} (R_{N+1};\eff,\textrm{h})= a_\nu^\text{P}\left(\lbrace R_1,\ldots,R_N\rbrace;\lbrace A_1,\ldots,A_N\rbrace,\textrm{h}\right)
\end{equation}
where $a_\nu^\text{P}(R_{N+1};\eff,\textrm{h})$ are the scattering coefficients of a single-layered cylinder with electric permittivity $\varepsilon_\eff$, magnetic permeability $\mu_\eff$ and radius $R_{N+1}$, embedded in a host with electric permittivity $\varepsilon_h$ and magnetic permeability $\mu_h$. 
In the right-hand-side of Eq.~\eqref{eq:generalized-cpa-index}, $a_\nu^\text{P}(\lbrace R_1,\ldots,R_N\rbrace;\lbrace A_1,\ldots,A_N\rbrace,\textrm{h})$ are the scattering coefficients of the original cylinder, consisting of $N$ co-centered layers with radii $\lbrace R_1,R_2,\ldots, R_N\rbrace$ and materials (including surface conductivities at each interface) $\lbrace A_1, A_2,\ldots, A_N\rbrace$ embedded in the host material (of the original system). 
In the metamaterial frequency range, $k_h R_{N+1}<1$, however, there are only two dominant modes per polarization, the lower order ones, i.e. the $\nu=0$ and the $\nu=1$ mode.
In the limit $k_{\text{eff}} R_{N+1} \ll 1$, we can replace the Bessel functions in Eq.~\eqref{eq:generalized-cpa-index}  with their limiting expressions assuming small arguments~\cite{Stegun}. Considering only the $\nu=0$ and $\nu=1$ modes per polarization, we can obtain semi-analytical expressions for all the components of the effective electric permittivity and magnetic permeability tensors, which read as
\begin{align}
\varepsilon_{\eff}^{\parallel} &= -\frac{2\varepsilon_h}{k_hR_{N+1} }\left[\frac{J_0'(k_hR_{N+1})+H_0'(k_hR_{N+1})a_0^{\TM}}{J_0(k_hR_{N+1})+H_0(k_hR_{N+1})a_0^{\TM}}\right] \label{eq:cpa1b}\\
 \mu_{\eff}^{\perp} &=   \frac{\mu_h}{k_hR_{N+1}} \left[ \frac{J_1(k_hR_{N+1}) + H_1(k_hR_{N+1})a_1^{\TM}}{J_1'(k_hR_{N+1})+H_1'(k_hR_{N+1})a_1^{\TM}} \right]\label{eq:cpa2b} \\
\mu_{\eff}^{\parallel} &= -\frac{2\mu_h}{k_{h}R_{N+1}} \left[ \frac{J_0'(k_hR_{N+1}) + H_0'(k_{h}R_{N+1})a_0^{\TE}}{J_0(k_{h}R_{N+1})+H_0(k_hR_{N+1})a_0^{\TE}} \right] \label{eq:cpa3b}
\\
\varepsilon_{\eff}^{\perp} &=  \frac{\varepsilon_{h}}{k_hR_{N+1}} \left[ \frac{J_1(k_{h}R_{N+1}) + H_1(k_hR_{N+1})a_1^{\TE}}{J'_1(k_hR_{N+1})+H'_1(k_hR_{N+1})a_1^{\TE}} \right] \label{eq:cpa4b}
\end{align}
where the coefficients $a_0^{\TE},a_1^{\TE},a_0^{\TM},a_1^{\TM}$ are the scattering coefficients of the original $N-$layer cylinder embedded in the host of the original system (the ones of the r.h.s. of Eq.~\eqref{eq:generalized-cpa-index}).
Equations \eqref{eq:cpa1b}-\eqref{eq:cpa4b} have the same form as the relations obtained in Ref.~\cite{Mavidis2020PRBpol}. 
Finally, if we further take the limit $k_hR_{N+1}\ll 1$, we get 
\begin{align}
\varepsilon_\eff^\parallel &= \varepsilon_h\left[1  -\frac{f}{(k_hR_N)^2} \frac{4i}{\pi}a_0^\TM \right] \label{eq:epspar-lim}\\
\mu_\eff^\perp &= \mu_h\left[\frac{(k_hR_N)^2-f\frac{4i}{\pi}a_1^\TM}{(k_hR_N)^2+f\frac{4i}{\pi}a_1^\TM} \right] \\
\mu_\eff^\parallel &= \mu_h\left[1  -\frac{f}{(k_hR_N)^2} \frac{4i}{\pi}a_0^\TE \right] \\
    \varepsilon_\eff^\perp &= \varepsilon_h\left[\frac{(k_hR_N)^2-f\frac{4i}{\pi}a_1^\TE}{(k_hR_N)^2+f\frac{4i}{\pi}a_1^\TE} \right] \label{eq:epsperp-lim}
\end{align}
where $f$ is the cylinders filling ratio in the system/metamaterial. 
Note here that for cylinders without any coating ($N=1$, $\sigma_e=0$, $\sigma_m=0$) in the quasistatic limit (i.e. $k_hR_{N+1}\ll1$, $k_1R\ll1$), Eqs.~\eqref{eq:cpa1b}-\eqref{eq:cpa4b} reduce to the well-known Maxwell-Garnett expressions.


\section{Results and Discussion}
\label{sec:metasurface-results}
Here we apply the the methods presented in the previous Section in systems of graphene cylinders/tubes and metasurface-made cylinders, which are representative systems for the demonstration of the potential of our approaches, as well as systems associated with novel and engineerable optical properties.

\subsection{Single scattering}
\label{subsc:results-singlesc} 
\subsubsection*{\em Graphene cylinders}

\begin{figure}[tb]
  \begin{center}
   \includegraphics[width=86mm]{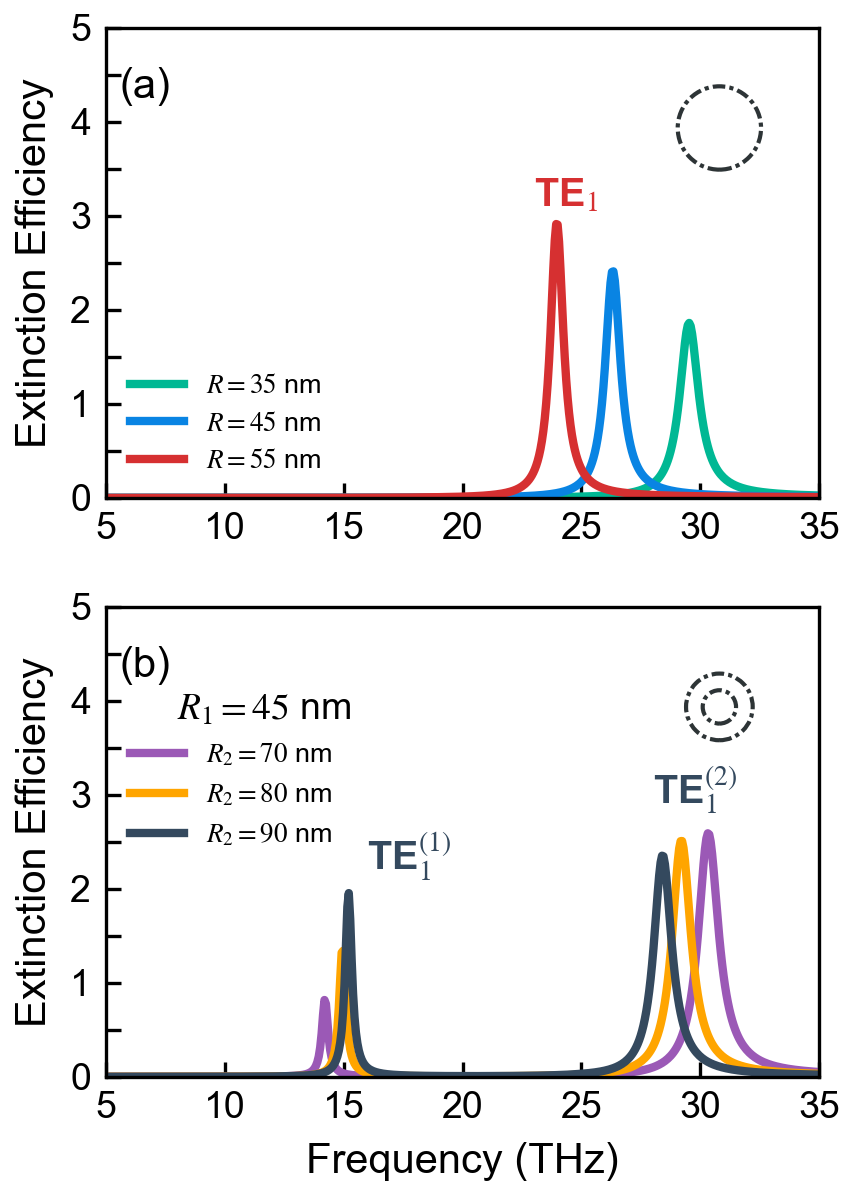}
    \caption{\label{fig:CNT-Qext} Extinction efficiencies, $Q_{\rm{ext}}$, in (a) a single-layer graphene cylindrical tube in air with varying radius, $R$, and (b) a double-layered graphene cylinder in air with inner radius $R_1=45$~nm, and variable outer layer radii $R_2$ for TE polarization . All layers exhibit surface  conductivity $\sigma_g$ with Fermi level $E_F=0.2$~eV and relaxation time $\tau=1$~ps. All cylinder bulk (inter-surface) layers are made of air. }

  \end{center}
\end{figure}

We begin our analysis by calculating the extinction efficiencies of (i) a single-layered and (ii) a double-layered cylinder, formed by homogeneous graphene layers.
 We assume that all the bulk cylinder layers are  air, i.e.  $\varepsilon_\ell=1$ and  $\mu_\ell = 1$, and the same for the host material. 
The geometry is comprised, in fact, of co-centered cylindrical cells/sheets with electric surface conductivity $\sigma_e=\sigma_g$ calculated using Eq.~\eqref{eq:graphene_rpa2} and plotted in Fig.~\ref{fig:graphene-cond}(a). (Such a geometry can be considered as a good approximation of a family of single- and double-wall carbon nanotubes.) 
The extinction efficiencies  for a  single graphene cylindrical layer of variable radius, $R=35$ nm, $R=45$ nm and $R=55$ nm, are shown in Fig.~\ref{fig:CNT-Qext}(a), while  the extinction efficiencies for a double-layered cylinder
 with variable outer-layer radius  $R_2$ are shown in Fig.~\ref{fig:CNT-Qext}(b).
For the single graphene layer  [Fig.~\ref{fig:CNT-Qext}(a)] there is only one dominant peak in the extinction spectrum, originated from the dipolar $\nu=1$  mode for TE polarization. We denote this mode as TE$_1$.
Since, in this frequency region the imaginary part of the surface conductivity of graphene [Fig.~\ref{fig:graphene-cond}(a)] is positive, the mode is similar in nature to the Localized Surface Plasmon Resonance (LSPR) sustained in metallic rods~\cite{Pfeiffer1974PRB}.

\begin{figure*}[t!]
\begin{center}
   \includegraphics[width=1\linewidth]{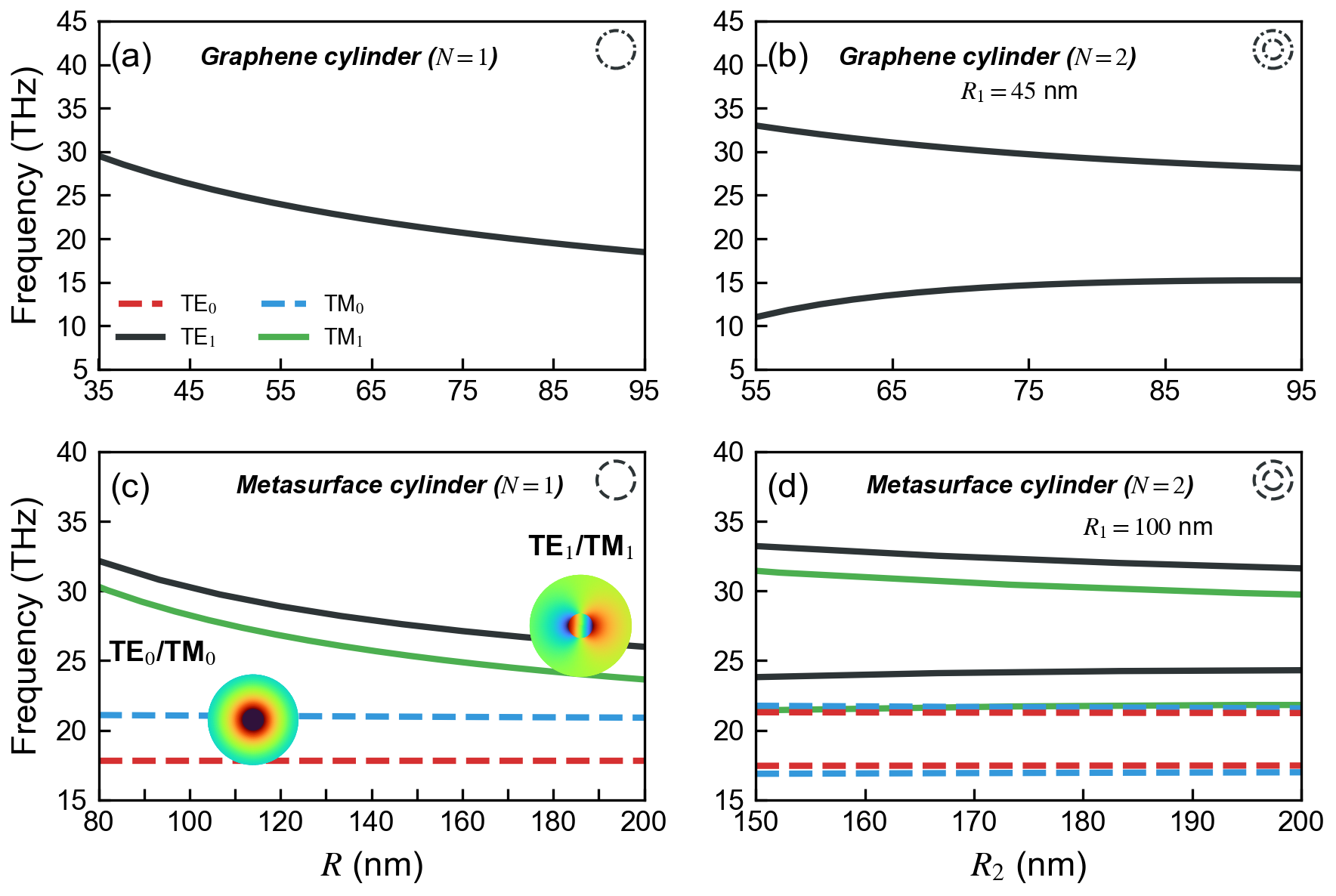}
    \caption{\label{fig:modes} Resonance frequencies of first two modes, $\nu=0$ and $\nu=1$,  per polarization, TE and TM,  for: (a) a single-layered graphene cylinder versus its radius, $R$; (b) a double-layered graphene cylinder for different outer-layer radii, $R_2$ and constant core radius $R_1=45$~nm ; (c) a single-layered metasurface-formed cylinder of different radii, $R$; (d) a double-layered metasurface cylinder with core radius $R_1=100$~nm and variable interlayer radii $R_2$. The characteristic field  distributions of the $z$ component of the electric (magnetic) field for TM (TE) polarization for the $\nu=0$ and $\nu=1$ modes are shown in the insets of (c). }
  \end{center}
\end{figure*}
 \begin{figure*}[t!]
\begin{center}
   \includegraphics[width=1\linewidth]{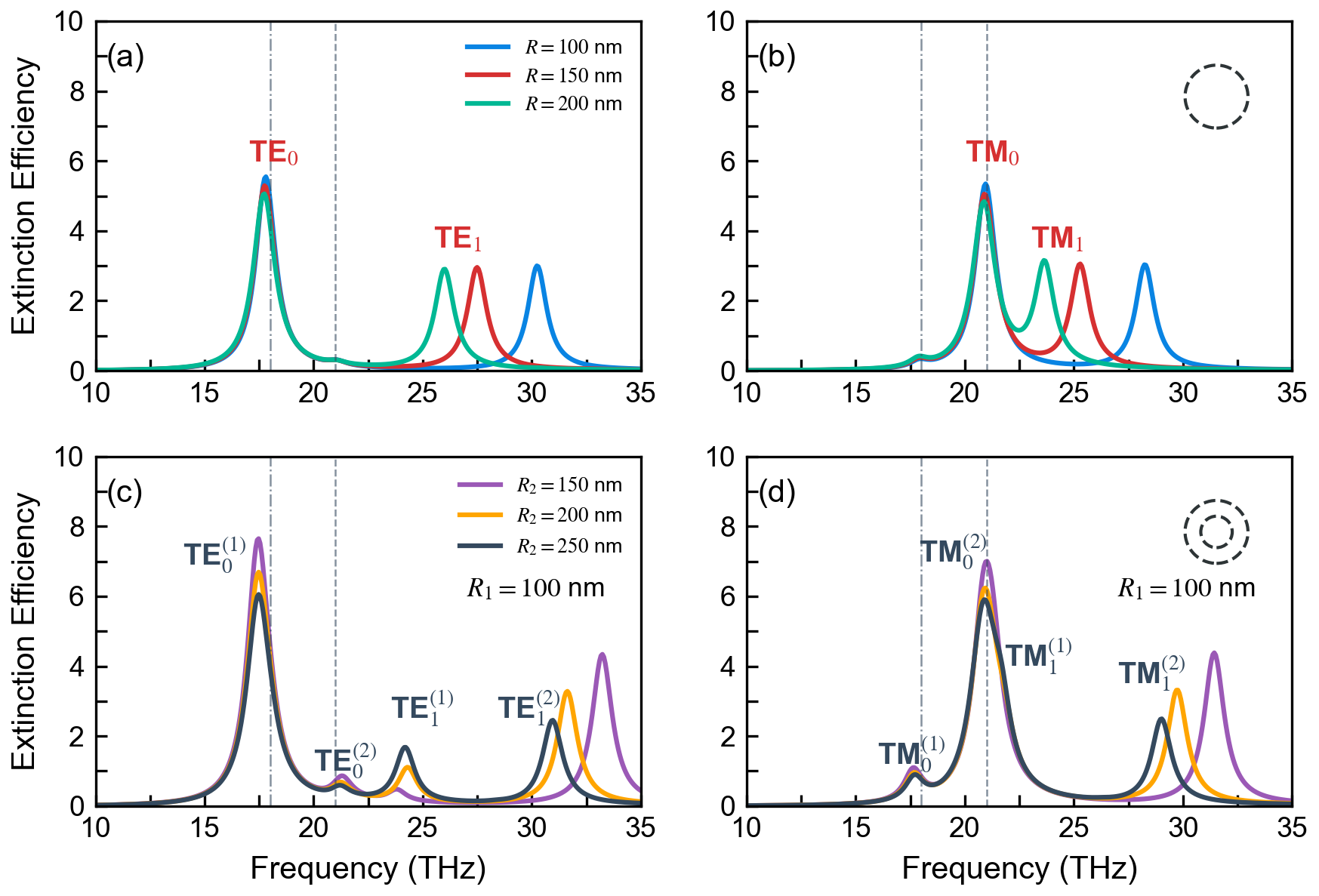}
    \caption{\label{fig:Huygens-Qext} Extinction efficiencies, $Q_{\rm{ext}}$, for:(a) and (b), a single-layered ($N=1$) metasurface-formed cylinder of variable radius, $R$, for TE (a) and TM (b) polarization;  (c) and (d), a double-layered ($N=2$)  metasurface cylinder with constant core radius $R_1=100$~nm and variable outer-layer radius, $R_2$, for TE (c) and TM (d) polarization. The vertical dashed grey lines in the panels indicate the resonant frequencies of the metasurface magnetic and the electric conductivities at $f_m=$18~THz and $f_e=$21~THz, respectively. The symbols associated with each resonance are explained in the main text.}
  \end{center}
\end{figure*}

For the double-layered  case ($N=2$) shown in  Fig.~\ref{fig:CNT-Qext}(b) we consider a core of constant radius $R_1=45$ nm and the variable outer layer  radius:  $R_2=70$ nm,  $R_2=80$ nm  and  $R_2=90$ nm.
In this case, the TE$_1$ mode manifests as two distinct resonances at frequencies below and above the TE$_1$ resonance for the single layer case. Here, we denote these resonances in order of increasing frequency as TE$_1^{(1)}$ and TE$_1^{(2)}$.
From the  frequency of the modes, one can conclude that the mode TE$_1^{(1)}$ comes predominately from the  outer layer, red-shifted due to the presence and interaction with the inner one, while TE$_1^{(2)}$ from the contribution of the core-layer, blue-shifted due to the interaction with the outer layer.
A worth-mentioning feature of Fig.~\ref{fig:CNT-Qext}(b) is the small frequency shift of the  TE$_1^{(1)}$ peak with the change of the radius $R_2$, compared, e.g., with the corresponding shift observed in Fig.~\ref{fig:CNT-Qext}(a). The cause of this difference is the presence of the inner layer and the coupling between the two layers, as will be discussed also in the next paragraph. 
We should note here that for these systems the extinction efficiency  for TM polarization is below $10^{-2}$, with no resonances for both cases, and it is not shown here. 

The dependence of the resonance frequencies on the radii of the graphene cylindrical layers, which are calculated by finding the poles of the scattering coefficients $a_\nu$ [Eq.~\eqref{eq:sc-coeff}], are shown in Figs.~\ref{fig:modes}(a)-(b).
We observe that the  resonance frequency in the single-layered cylinder scales with the radius of the cylinder with a $1/\sqrt{R}$ dependence as we derived in Eq.~\eqref{eq:te1-lim} (see Sec.~\ref{subsec:metasurface-methods-singlescatt}).
The tendency of TE$_1$ modes for the double-layered case ($N=2$) can be explained in terms of mode hybridization and level-repulsion, where the modes coupling leads to a  lower-frequency ("\emph{bonding}") mode, red-shifted in respect to the "parent" single-layer mode (with the shifting being larger for smaller difference between $R_1$ and $R_2$), and a higher-frequency (\emph{anti-bonding}) mode, blue-shifted in respect to the corresponding single-layer mode~\cite{Prodan2003Science,Liu2015SciRep,Li2017IEEE, Raad2019JPD}. Comparing the  results of Figs.~\ref{fig:modes}(a) and (b) we observe that the mode shifting due to the interaction of the inner and outer graphene layer is quite significant. 

\begin{figure}[hhh]
  \begin{center}
   \includegraphics[width=86mm]{figures/Huygens_a0tm0_vs_fm_50nmB_nov.png}
    \caption{\label{fig:Huygens-a0} Zeroth-order scattering coefficients, $a_0$, as a function of the magnetic surface conductivity resonance frequency $f_m$ for (a) TE and (b) TM polarizations for 
    a double-layered metasurface cylinder ($N=2$) with radii $R_1=100$~nm and $R_2=150$~nm, and constant electric surface conductivity resonance frequency $f_e=21$~THz (dashed vertical line).
    }

  \end{center}
\end{figure}

\subsubsection*{\em Metasurface cylinders}

Next, we turn our attention to cylinders/tubes formed by metasurfaces having both electric and magnetic response, embedded in air.
We consider two cases: (i)  a single-layered metasurface-made cylinder of variable radius and (ii) a double-layered cylinder, of the same metasurface at each layer, with fixed inner-layer radius. The metasurfaces have resonant conductivities, $\sigma_{e(1)}=\sigma_{e(2)}$ and $\sigma_{m(1)}=\sigma_{m(2)}$ as it is  shown in Fig.~\ref{fig:graphene-cond}(b), and $\varepsilon_1=\varepsilon_2=\varepsilon_h=1$. 
The extinction efficiencies for both TE and TM polarizations and for the single- and the double-layered cylinders are shown in Fig.~\ref{fig:Huygens-Qext}. 
In  the single-layered cylinders we investigate cases of  radii $R=50$~nm, $R=100$~nm and $R=150$ nm.  There are two dominant resonances for both TE and TM polarizations. 
The electric in nature TM$_0$ and the magnetic in nature TE$_0$ modes appear just below the resonance frequencies of the electric sheet conductivity, $f_e=\omega_e/2\pi=21$~THz, and the magnetic sheet conductivity, $f_m=\omega_m/2\pi=18$~THz, respectively, and are practically independent of the cylinder radius. 
Just below $f_e$ ($f_m$) the imaginary part of the electric (magnetic) sheet conductivity of the metasurface is negative [see Fig.~\ref{fig:graphene-cond}(b)]  leading to positive equivalent electrical permittivity (magnetic permeability); this case is similar to the polaritonic cylinders for small radii we have discussed in Ref.~\cite{Mavidis2020PRBpol}.
On the other hand, the dipole-like modes, TE$_1$ and TM$_1$, fall in the regions of positive imaginary part of electric and magnetic conductivity respectively and they are similar in nature with the modes discussed earlier for the graphene case.

Next we study the case of the double-layered cylinder formed of the same metasurface, at inner layer radius $R_1=100$~nm and different outer layer radii $R_2$.  
We observe two dipolar ($\nu=1$) electric and magnetic modes, TE$_1^{(1)}$ and  TE$_1^{(2)}$, TM$_1^{(1)}$ and TM$_1^{(2)}$,   one of lower and one of higher frequency than the corresponding modes of a single-layered cylinder, similar in nature and behavior to the graphene case we discussed in the previous paragraph. 
Note here that the TM$_1^{(1)}$ mode [see Fig.~\ref{fig:modes}(d)] is very close in frequency with the TM$_0^{(2)}$ mode. Besides, there are the monopolar, $\nu=0$, modes which are practically the same with the single-layer case, with main difference the enhancement of the extinction of the secondary peaks $\TE_0^{(2)}, \TM_0^{(2)}$, coinciding also with conductivities' resonances.

The dependence of the resonance frequencies of the dominant modes on the radius of the single-layered cylinder and on the  radius $R_2$ for the double-layered case, for constant $R_1=100$~nm, is shown in Figs.~\ref{fig:modes}(c)-(d). 
We observe that the $\nu=1$ modes show similar behavior with the correspondent modes of the graphene case. The $\nu=0$ modes on the other hand, where their originating field (electric for TM and magnetic for TE) does not encounter material discontinuities along its direction, almost coincide in frequency with the  resonances of the conductivities and are unaffected by the cylinder radius.

It is interesting to observe though in the double-layer case the enhancement of the $\TE_0^{(2)}, \TM_0^{(2)}$ peaks compared to the single-layer case (where the peaks are hardly visible).  
To further elucidate this enhancement and the origin of those peaks we plot the zeroth-th order scattering coefficients $a_0$, in Fig.~\ref{fig:Huygens-a0}, for a double-layered cylinder with $R_1=100$~nm and $R_2=150$~nm as a function of the  magnetic sheet conductivity resonance frequency, $f_m$, with constant $f_e=21$~THz.
For instance, for the TE polarization, we can observe that the strength of  the weaker TE$_0^{(2)}$  resonance becomes more prominent and approaches in frequency the TE$_0^{(1)}$  as the resonance frequency $f_m$ approaches $f_e$.   Interestingly, even in the case of $f_e = f_m$, the two resonances remain distinct as a result of the  interaction of the layers about the electric and magnetic conductivity resonance. 
However, the secondary TE$_0$ mode is absent when the electric conductivity is zero (see Fig.~\ref{fig:Qext-2L-ZeroCond} in Appendix~\ref{app:Zero-conductivity}), indicating that this dip corresponds to a magnetic mode originating from the resonant electric conductivity response, in analogy with the resonant magnetic response obtained in polaritonic or high-index dielectric cylinders, but of much smaller strength. 
Analogous behaviour is observed in the case of the secondary TM$_0$ mode, which corresponds to electric response originated from large magnetism. 

\begin{figure*}[hhh]
  \begin{center}
   \includegraphics[scale=1]{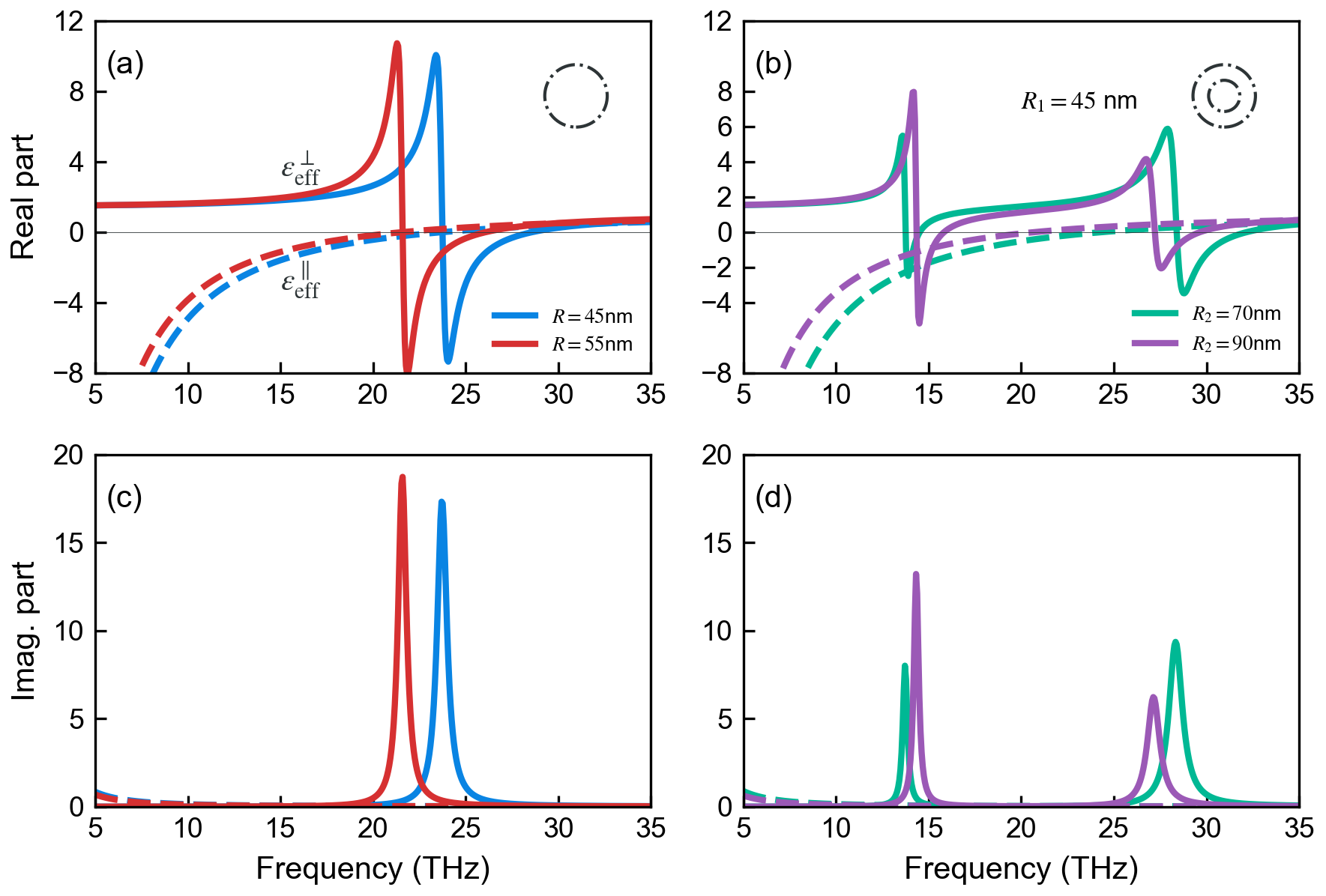}
    \caption{\label{fig:CNT-eff} Real (top row) and imaginary (bottom row) parts of the parallel $ \varepsilon_\eff^\parallel$ (dashed curves) and perpendicular $\varepsilon_\eff^\perp$  (solid curves) components of the relative effective permittivity $\varepsilon_\eff$  for: (a) and (c) a system made of single-layered ($N=1$) graphene cylinders of variable radius, $R=45$~nm (blue curves) and  $R=55$~nm (red curves) in air; (b) and (d) a system of double-layered ($N=2$) graphene cylinders in air, with constant core radius $R_1=45$~nm,   and for outer layer radii $R_2=70$~nm (green curves)   and  $R_2=90$~nm (purple curves); the cylinders filling ratio is in all cases equal to  $f=20\%$. }
  \end{center}
\end{figure*}

\begin{figure}[tb]
  \begin{center}
   \includegraphics[width=86mm]{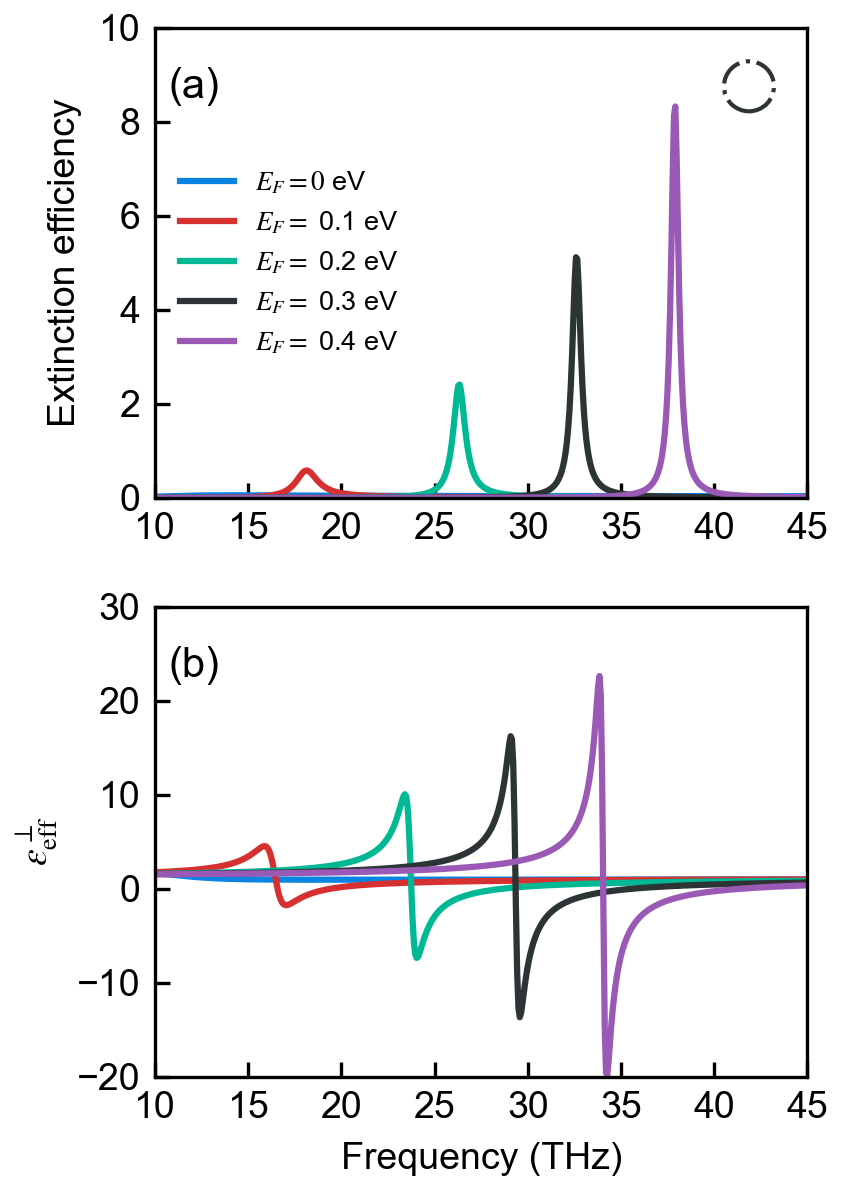}
    \caption{\label{fig:graphene1L-vsEf} (a) Extinction efficiency for TE polarization for  a single-layered graphene cylinder with radius $R=45$~nm in air, for different Fermi energies, $E_F$.  (b) Real part of the in-plane component of effective electric permittivity $\varepsilon_\eff^\perp$ for a system of single-layered graphene cylinders as the one of panel (a), in air, with cylinder filling ratio $f=$20\%.}

  \end{center}

\end{figure}

\begin{figure}[tbp]
  \begin{center}
   \includegraphics[width=86mm]{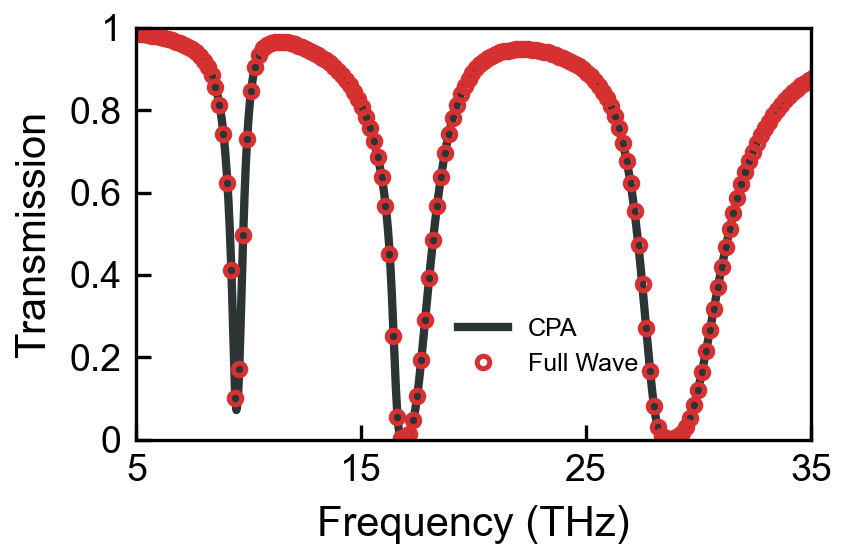}
    \caption{\label{fig:EMTverification}  Normal incidence, TE transmission spectra for a slab of
triple-layered graphene cylinders with radii $R_1=45$~nm, $R_2=90$~nm and $R_3=135$~nm and filling ratio $f = 20\%$
in air, in a square arrangement. The transmission is calculated
by the commercial finite element method electromagnetic solver
Comsol Multiphysics, considering a computational system of 7
unit cell thickness (along the propagation direction). The full wave transmission results (red circles) are compared with results for a homogeneous effective medium of the same thickness as the actual
system and effective parameters obtained through our CPA-based effective medium approach (black line).  }

  \end{center}
\end{figure}

\subsection{Effective medium theory and alternating optical phases}
\label{subsc:results-emt} 
\subsubsection*{{\em Graphene cylinders}}
We turn our attention now to the effective medium parameters $\varepsilon_\eff$ and $\mu_\eff$ for clusters of graphene and metasurface  cylinders in air, aligned in a square lattice, as presented in Fig.~\ref{fig:cylinder}(c). The cylinders considered are the ones discussed in the previous section. We examine initially the graphene cases. 
In Fig.~\ref{fig:CNT-eff} we plot the real and imaginary parts of the tensor components of effective electric permittivity $\varepsilon_\eff=\varepsilon^\perp_\eff(\hat{x}\hat{x}+\hat{y}\hat{y}) + \varepsilon_\eff^\parallel\hat{z}\hat{z}$ for single-layered graphene cylinders [Fig.~\ref{fig:CNT-eff}(a) and Fig.~\ref{fig:CNT-eff}(c)] and double-layered cylinders  [Fig.~\ref{fig:CNT-eff}(b) and Fig.~\ref{fig:CNT-eff}(d)] for constant filling ratio, $f=$20\%, and different radii.
As one can see, for both the single- and the double-layered cases there are lorentzian-shaped resonances for the in-plane component of the effective electric permittivity, $\varepsilon_\eff^\perp$, close in frequency to the corresponding TE$_1$ resonances [see  Figs.~\ref{fig:CNT-Qext} and \ref{fig:modes}(a)-(b)].
Also we can observe a Drude-like response for the parallel component of effective electric permittivity $\varepsilon_\eff^\parallel$ for both arrays of single- and double-layered graphene cylinders.
All the components of the effective magnetic permeability $\mu_\eff$ are equal to unity and are not shown here. 
As the radius of the cylinders increases the resonance of the $\varepsilon_\eff^\perp$ moves to lower frequencies.
Interestingly, for single-layered cylinders all the components of the effective electric permittivity vanish at the same frequency, close to the TE$_1$ mode resonance frequency. 
For instance, for $R=45$~nm the Epsilon-Near-Zero (ENZ) is achieved at  23.7~THz and moves to lower frequencies as the radius of the cylinder increases. 
This result along with Eqs.~\eqref{eq:cpa1b} and \eqref{eq:cpa4b} suggests that $a^\TM_0=a^\TE_1$ at that frequency for the single-layered graphene cylinder.
Further examinations showed that the monochromatic vanishing of both components of effective permittivity tensor holds only for the symmetric case, where the material inside the graphene layer and the host are the same (air here). 
On the other hand, for the double-layered cylinders, due to the presence of multiple resonances, there are several frequencies where the $\varepsilon_\eff^\perp$ vanishes. 

Moreover, as one can notice, there are frequency regions where the in-plane and out-of-plane components of the effective electric permittivity have different signs, i.e. $\varepsilon_\eff^\perp \cdot \varepsilon_\eff^\parallel<0$. This is the condition for hyperbolic response (i.e. dispersion relation of the shape of hyperbola) for TM-polarized waves, as can be seen by taking into account  the dispersion relations for an anisotropic homogeneous material~\cite{Depine2006JOSA},
\begin{align} \label{eq::hmm-tm}
    \textrm{TM:} \: \: \frac{k_\perp^2}{\mu_\eff^\perp\varepsilon_\eff^\parallel} + \frac{k_\parallel^2}{\mu_\eff^\perp\varepsilon_\eff^\perp} = \left(\frac{\omega}{c}\right)^2, \\
    \textrm{TE:} \: \: \frac{k_\perp^2}{\mu_\eff^\parallel\varepsilon_\eff^\perp} + \frac{k_\parallel^2}{\mu_\eff^\perp\varepsilon_\eff^\perp} = \left(\frac{\omega}{c}\right)^2, \label{eq::hmm-te}
\end{align}
where $k_\parallel$ and $k_\perp$ stand for the wave-vector components parallel and perpendicular to the cylinder axis respectively.
For example, for the single-layered graphene cylinders with radius $R=45$ nm [blue curve in Fig.~\ref{fig:CNT-eff}(a)] the condition for hyperbolic response is achieved up to 28.6~THz, where both $\varepsilon_\eff^\perp$ and $\varepsilon_\eff^\parallel$ become positive. 
We can further distinguish the hyperbolic metamaterial response of our systems into two different frequency regions by considering the different signs of $\varepsilon_\eff^\perp$ and $\varepsilon_\eff^\parallel$. 
For frequencies below the TE$_1$ resonance (26.7 THz) the in-plane components $\varepsilon_\eff^\perp$ are positive, while the out-of-plane component $\varepsilon_\eff^\parallel$ is negative; thus we have  \emph{type I hyperbolic metamaterial (HMM I)}. 
On the other hand, in the frequency region 26.7~THz-28.6~THz  we have $\varepsilon_\eff^\perp<0$ and $\varepsilon_\eff^\parallel>0$, thus   \emph{hyperbolic metamaterial type II (HMM II)} response.
The response is  more rich  for the metamaterial comprised of double-layered graphene cylinders shown in  Fig.~\ref{fig:CNT-eff}(b) and Fig.~\ref{fig:CNT-eff}(d).
Considering the case with constant core radius, $R_1=45$ nm, and variable outer layer radii, $R_2=70$ nm [green curve in Fig.~\ref{fig:CNT-eff}(b) and Fig.~\ref{fig:CNT-eff}(d)]  and $R_2=90$ nm [purple curve in Fig.~\ref{fig:CNT-eff}(b) and Fig.~\ref{fig:CNT-eff}(d)], we see that there are alternating optical phases (HMM I, metallic, HMM II and dielectric) at frequencies close to the two TE$_1$ resonances. For instance, for $R_2=70$~nm we find HMM I response in the frequency region up to 13.7~THz and 14.6-24.3~THz and HMM II response in the frequency region 28.5 - 32.~THz. 

We should mention also here that the response of graphene-shells and graphene-coated cylinders is highly tunable by changing the Fermi level of graphene. This, as mentioned, can be accomplished by various methods, including chemical doping, voltage tuning and photoexcitation. To assess the effect of this tunability on the effective properties of our graphene-based metamaterial we calculate first the dependence of the extinction efficiency on the graphene Fermi level, $E_F$, for a single-layered graphene cylinder - see  Fig.~\ref{fig:graphene1L-vsEf}(a).  As observed in Fig.~\ref{fig:graphene1L-vsEf}(a), the extinction efficiency is very small (maximum of $Q_{\textrm{ext}}=0.05$) for the case of zero Fermi level; as the Fermi energy increases the extinction efficiency becomes larger and the resonance shifts to higher frequencies.
Analogous trends are observed for the in-plane component of the effective electric permittivity $\varepsilon_\eff^\perp$, which is shown in Fig.~\ref{fig:graphene1L-vsEf}(b).  Fig.~\ref{fig:graphene1L-vsEf}(b) shows $\varepsilon_\eff^\perp$ for  cylinders with radius $R=45$~nm and filling ratio $f=$20\%, for different Fermi energies, $E_F$=0, 0.1, 0.2, 0.3, 0.4 eV. The effective permittivity results are presented in parallel with single-scattering data (Fig.~\ref{fig:graphene1L-vsEf}(a)), to facilitate the understanding of the observed response. 
We see that the TE$_1$ resonance  of the single scattering setup [Fig.~\ref{fig:graphene1L-vsEf}(a)]  is moving towards higher frequencies as the Fermi level grows, and the resonance becomes stronger in both extinction efficiency and effective electric permittivity spectra.

\subsubsection*{{\em CPA accuracy}}
 In order to verify and demonstrate the validity and accuracy of the developed effective medium approach we compare its results with equivalent full-wave simulations data. Specifically we calculate the transmission and reflection spectra through a slab consisting  of seven unit cells (along propagation direction) of a triple-layered ($N=3$) graphene cylinder,  in  square arrangement, using the full wave numerical analysis software COMSOL MULTIPHYSICS, and we compare the results with the response (obtained by transfer matrix calculations) of a slab of the same thickness with electric permittivity $\varepsilon_\eff$ and magnetic permeability $\mu_\eff$ calculated through our developed formalism. 
The transmission comparison for TE polarization is shown in Fig.~\ref{fig:EMTverification}.
As can be seen, there is an excellent agreement between full wave simulations and the effective medium model. Moreover, in this particular system, the subwavelength size of the graphene tubes allows an accurate description by the CPA for frequencies even higher than the third structure resonance.  

\subsubsection*{{\em Metasurface-based cylinders}}

\begin{figure*}[hhh]
  \begin{center}
   \includegraphics{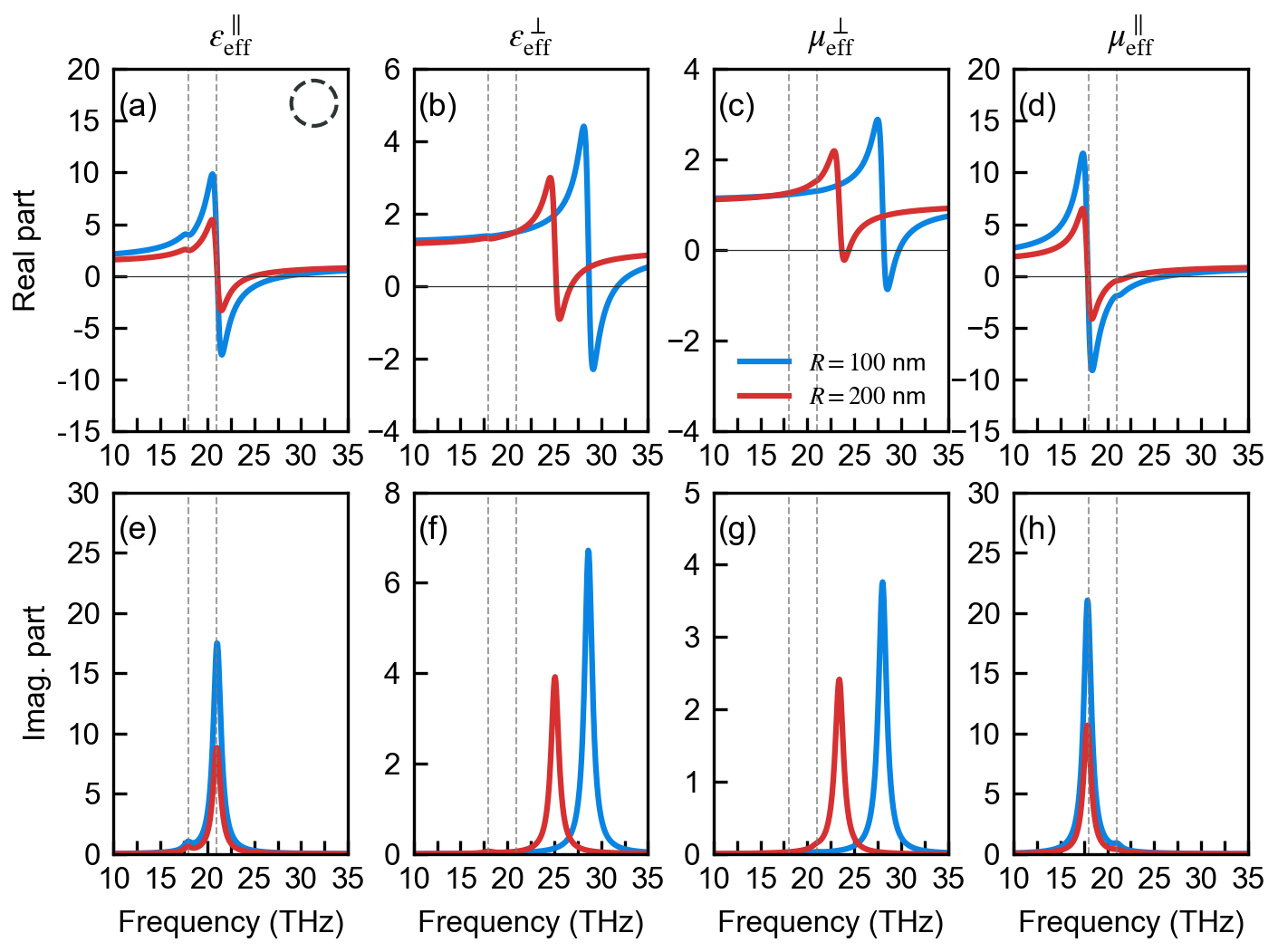}
    \caption{\label{fig:ElMagn-Effective-1L}
Real (top row) and imaginary  (bottom) parts of the tensor components of effective  permittivity $\varepsilon_\eff$ (first column  $\varepsilon_\eff^\parallel$   and second column  $\varepsilon_\eff^\perp$) and effective  permeability $\mu_\eff$ (third column  $\mu_\eff^\perp$   and fourth column  $\mu_\eff^\parallel$) for a system/metamaterial of single-layered ($N=1$) metasurface-formed cylinders in air, for different radii, $R=100$~nm (blue lines) and $R=200$~nm (red lines),  and filling ratio $f=20$\%. The vertical dashed  lines indicate the resonance frequencies of the magnetic ($f_m=18$~THz) and electric sheet conductivity ($f_e=21$~THz). 
}
  \end{center}
\end{figure*}

\begin{figure*}[hhh]
  \begin{center}
   \includegraphics{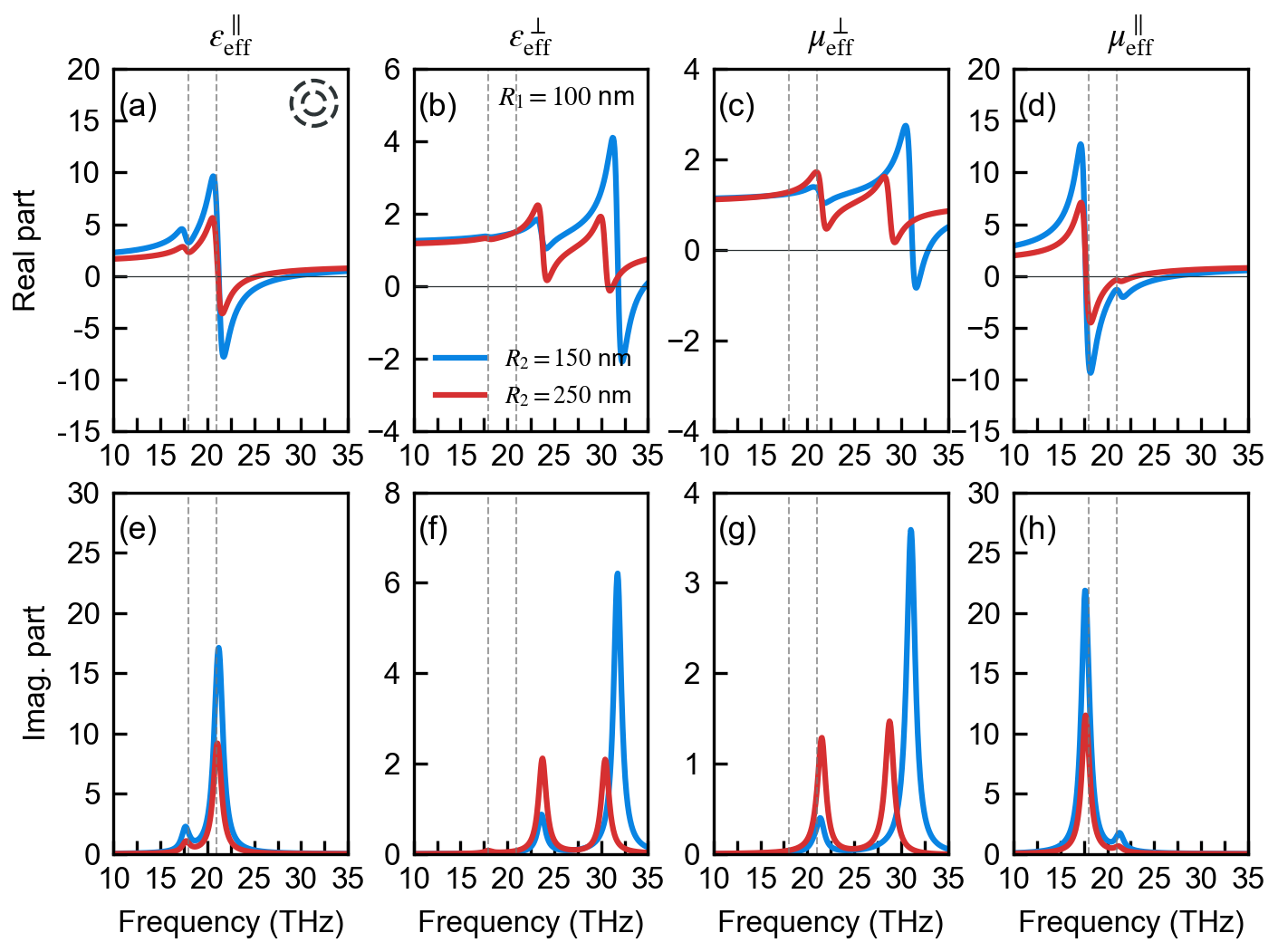}
    \caption{\label{fig:ElMagn-Effective-2L}
Real (top row) and imaginary (bottom row)   parts of the tensor components of effective  permittivity $\varepsilon_\eff$  (first column  $\varepsilon_\eff^\parallel$,  second column  $\varepsilon_\eff^\perp$) and effective  permeability $\mu_\eff$ (third column  $\mu_\eff^\perp$, fourth column  $\mu_\eff^\parallel$) for a system/metamaterial of double-layered ($N=2$) metasurface-formed cylinders in air, with core radius $R_1=100$~nm and for different outer-layer radii, $R_2=150$~nm (blue lines) and  $R_2=250$~nm (red lines), and filling ratio $f=20$\%. The vertical dashed lines indicate the resonance frequencies of the magnetic ($f_m=18$~THz) and electric sheet conductivity ($f_e=21$~THz). 
}
  \end{center}
\end{figure*}

Next, we turn our attention to the effective response of arrays of metasurface-based cylinders, investigating the single-layered and double-layered cylinder systems as previously; the results are presented in Fig.~\ref{fig:ElMagn-Effective-1L}  and  Fig.~\ref{fig:ElMagn-Effective-2L}, respectively. 
The effective medium parameters for a single-layered cylinder metasurface  for both TE and TM polarizations are shown in Fig.~\ref{fig:ElMagn-Effective-1L} for two cylinders radii, $R=100$~nm (blue curves) and $R=200$~nm (red curves). 
As one can notice in Fig.~\ref{fig:ElMagn-Effective-1L}, the parallel components $\varepsilon_\eff^\parallel$ and $\mu_\eff^\parallel$ (originating from the zero-th order modes) become resonant at the electric sheet conductivity and magnetic conductivity resonances of the constituent metasurfaces, respectively, while the change of radius only affects the strength of the resonant response. On the other hand, the perpendicular components, $\varepsilon_\eff^\perp$ and $\mu_\eff^\perp$, strongly depend, in both resonance frequency and strength, on the cylinder radius. The dependence of the resonance frequency follows the corresponding dependence of their "parent" single scattering modes TE$_1$, TM$_1$ (see Figs. 5 (a) and (b))  while the strength is favored from smaller radii. 

Regarding the achievable metamaterial-related possibilities,  we can observe that 
 there are  frequency regions where the medium becomes \emph{double negative} (DNG) resulting in negative refractive index, i.e. both $\varepsilon_\eff<0$ and its pertinent $\mu_\eff<0$. 
For $R=100$~nm and TM polarization both $\varepsilon_\eff^\parallel$ and $\mu_\eff^\perp$ are negative between frequencies 28.19~THz and 28.64~THz.
For the larger cylinder-radius system DNG is achieved for TM polarization, at  lower frequencies, while 
for TE polarization there is no DNG response ($\varepsilon_\eff^\perp<0$ and $\mu_\eff^\parallel<0$) for the parameters studied here.

It is interesting to observe also that the metasurface-cylinders system exhibits HMM response for both TM and TE polarizations. 
By considering the anisotropic material dispersion relations of  Eqs.~\eqref{eq::hmm-tm}-\eqref{eq::hmm-te} and the results shown in Fig.~\ref{fig:ElMagn-Effective-1L} one can see that there are both HMM I and HMM II regions for both TM and TE polarizations. 
In particular, the arrays of metasurface-coated cylinders with $R=100$~nm (blue curves in Fig.~\ref{fig:ElMagn-Effective-1L}) and filling ratio $f=20$\% exhibit HMM I response ($\mu_\eff^\perp \varepsilon_\eff^\parallel<0$ and $\mu_\eff^\perp \varepsilon_\eff^\perp>0$) in the frequency ranges 21.1-28.15~THz and 28.7-29.75~THz, and HMM II ($\mu_\eff^\perp \varepsilon_\eff^\parallel>0$ and $\mu_\eff^\perp \varepsilon_\eff^\perp<0$) response in the frequency ranges 28.2-28.6~THz and 29.8-31.7~THz for TM polarization.
Also, for TE polarization there is  HMM I response ($\varepsilon_\eff^\perp \mu_\eff^\parallel<0$ and $\varepsilon_\eff^\perp \mu_\eff^\perp>0$) in the frequency regions 17.97-26.53~THz and 28.75-29.75~THz, and HMM II response ($\varepsilon_\eff^\perp \mu_\eff^\parallel>0$ and $\varepsilon_\eff^\perp \mu_\eff^\perp<0$)  in the frequency range 28.19-28.7~THz.

We close our analysis by investigating  the effective medium parameters for the  double-layered metasurface cylinders systems, which are shown  Fig.~\ref{fig:ElMagn-Effective-2L}.
We observe that for the parallel components $\varepsilon_\eff^\parallel$ and $\mu_\eff^\parallel$ the presence of the second (outer) layer does not affect the resonance position. Regarding the resonance strength the outer layer seems to rather dominate or screen the response of the inner one. For the perpendicular components though we observe addition of resonances and "repulsion" of modes and in a way analogous to what was observed in the graphene case. 
As in the single-layered case we observe also here a rich electromagnetic response, with  regions of alternating optical phase (between HMM I and HMM II) for both TE and TM polarizations. Regarding DNG response, although it is not observed in the results of  Fig.~\ref{fig:ElMagn-Effective-2L},  our calculations suggest that it can be achieved also for the double-layered case by properly tuning the structure parameters (radii, filling ratio).

\section{Conclusions}
\label{sec:conclusions} 
In this work we derived analytically  single scattering cross-sections and effective medium formulas for systems of  multi-layer co-centric cylinders.  Every layer can be made  of any material and can be coated with a metasurface of arbitrary resonant sheet-conductivities, both electric and magnetic. Starting from the investigation of the single cylinder’s scattering properties and resonances, for the formulation we combined Mie theory with a transfer matrix approach for cylindrical waves. The effective medium  derivation was based on the Coherent Potential Approximation method and provided semi-analytical expressions for the calculated effective medium parameters. Compared to commonly used effective medium approaches, the develop formulation is suitable also quite beyond the long-wavelength limit, is able to describe metamaterials made of resonant materials, predicts magnetism in all-dielectric media and accommodates cylindrical systems coated with resonant electromagnetic sheets.

We applied the formalism  into two different metamaterial systems, operating in the technologically appealing THz region: (a) graphene nanotubes of one and two concentric layers (approximating systems of carbon nanotubes) and variable conductivity, and (b)  nanotubes formed from a metasurface having both electric and magnetic response. We found that by properly choosing the number of layers and radii, both systems can exhibit a rich palette of   electromagnetic response, own to the  engineerable permittivity and permeability; this rich response includes   hyperbolic behavior of both type I and type II  for both TE and TM polarizations, double negative response and regions of epsilon-near-zero and mu-near-zero response. Thus, our results suggest that multi-layer  and metasurface-coated cylinders can be exploited for the design of multifunctional metamaterials and devices for the control of electromagnetic radiation, offering a vast range of possibilities, from superscattering to cloaking and advanced wavefront manipulation. 

--------------------------------------------
\section*{ACKNOWLEDGEMENTS}
We acknowledge financial support by  the European Union’s Horizon
2020 FETOPEN programme under projects VISORSURF (grant agreement No.
736876), NANOPOLY (grant agreement No. 829061) and SMARTWAVE (grant agreement No. 952088)  and by the
General Secretariat for Research and Technology, and the
H.F.R.I. Ph.D. Fellowship Grant (Grant Agreement No. 4894)
in the context of the action “1st Proclamation of Scholarships
from ELIDEK for Ph.D. Candidates.”
\appendix
\begin{figure}[ht]
  \begin{center}
   \includegraphics[width=86mm]{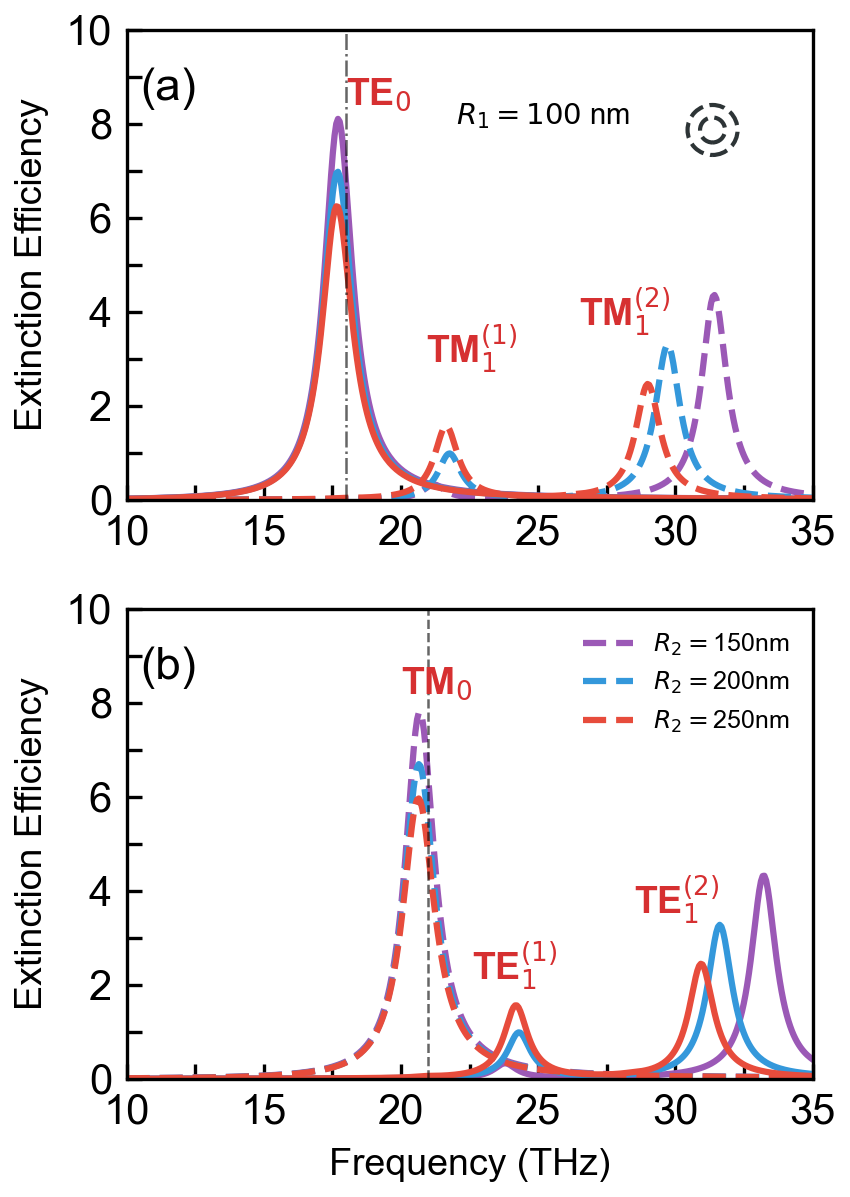}
    \caption{\label{fig:Qext-2L-ZeroCond} TE (solid lines) and TM (dashed lines) extinction efficiencies for a double-layered metasurface-coated cylinder (N=2) with core radius $R_1=100$~nm and various interlayer distances $\Delta_2=R_2-R_1$ for   metasurfaces with  (a) magnetic surface conductivity $\sigma_m$ with $f_m=18$~THz and $\sigma_e=0$, and (b) electric surface conductity $\sigma_e$ with $f_e=21$~THz and $\sigma_m=0$. The vertical dashed grey lines indicate the resonances of the sheet conductivities.  }
  \end{center}
\end{figure}

\section{Appendices}
\subsection{Transfer Matrix Method}
\label{app1:tmm}
In this appendix we derive the  transfer matrices used in our analysis.
For TE polarization  the magnetic field is parallel to the cylinder axis, \textbf{H}=$H_z\hat{z}$. 
The electric and magnetic fields in the $\ell$-th layer can be expanded on the basis of the cylindrical vector harmonics as 
\begin{equation}
\label{eq:y-El}
 \mathbf{E}_\ell = i\sum_{\nu=-\infty}^{\infty} \frac{i^\nu}{k_\ell}\left[c_{\ell\nu} \mathbf{M}^{(\text{outward})}_{e\nu k_\ell} + d_{\ell \nu} \mathbf{M}^{(\text{inward})}_{e\nu k_\ell} \right]
\end{equation}
\begin{equation}
\label{eq:Hz}
 \mathbf{H}_\ell = \frac{1}{\eta_\ell}\sum_{\nu=-\infty}^{\infty} \frac{i^\nu}{k_\ell} \left[ c_{\ell\nu} \mathbf{N}_{e\nu k_\ell}^{(\text{outward})} + d_{\ell\nu} \mathbf{N}_{e\nu k_\ell}^{(\text{inward})} \right]
\end{equation}
where $c$ is the speed of light in vacuum, $k_\ell = \sqrt{\varepsilon_\ell \mu_\ell} \frac{\omega}{c}$ and $\eta_\ell = \sqrt{\mu_\ell/\varepsilon_\ell}$.
The coefficients $c_{\ell\nu}$ and $d_{\ell\nu}$ can be determined from the boundary conditions at the surface of the cylinder.

The boundary conditions at the interface between the $\ell$-th and the $(\ell+1)$-th layer (i.e. at $r=R_{\ell}$) can be written as
\begin{equation}
\label{eq:bc1b} 
E^{\varphi}_{\ell+1} -  E^{\varphi}_\ell = -\sigma_{m(\ell)} \frac{H^{z}_\ell + H^{z}_{\ell+1}}{2}
\end{equation}
\begin{equation}
\label{eq:bc2b} 
H^{z}_{\ell+1} - H^{z}_\ell = -\sigma_{e(\ell)} \frac{E^{\varphi}_\ell + E^{\varphi}_{\ell+1}}{2}
\end{equation}
Using these conditions we can connect the fields in the $\ell$-th and $(\ell+1)$-th layer of the multi-layer cylinder with a transfer matrix as 
\begin{equation}
\mathbb{T}_{\ell \nu}^\text{P}
\begin{pmatrix}
d_{\ell \nu} \\[6pt]
c_{\ell \nu}
\end{pmatrix} = 
\begin{pmatrix}
d_{(\ell+1),\nu} \\[6pt]
c_{(\ell+1),\nu}
\end{pmatrix}
\end{equation}
where $\mathbb{T}_{\ell \nu}^\text{P}$ is the transfer matrix  for polarization $\text{P}=\text{TE}$.

For $\text{P}=\text{TM}$ polarization (the electric field is parallel to the cylinder axis, \textbf{E}=$E_z\hat{z}$) the electric and magnetic fields in the $\ell$-th layer can be expanded  as 
\begin{equation}
\label{eq:y-Hl}
 \mathbf{H}_\ell = -\frac{i}{ \eta_\ell}\sum_{\nu=-\infty}^{\infty} \frac{i^\nu}{k_\ell} \left[c_{\ell\nu} \mathbf{M}^{(\text{outward})}_{e\nu k_\ell} + d_{\ell\nu} \mathbf{M}^{(\text{inward})}_{e\nu k_\ell} \right]
\end{equation}
\begin{equation}
\label{eq:Ez}
 \mathbf{E}_\ell = \sum_{\nu=-\infty}^{\infty} \frac{i^\nu}{k_\ell}  \left[ c_{\ell\nu} \mathbf{N}_{e\nu k_\ell}^{(\text{outward})} + d_{\ell\nu} \mathbf{N}_{e\nu k_\ell}^{(\text{inward})} \right]
\end{equation}
The boundary conditions between the $\ell$-th and the $(\ell+1)$-th layer can be written as:
\begin{equation}
\label{eq:bc1b-tm} 
E^{z}_{\ell+1} -  E^{z}_\ell = \sigma_{m(\ell)} \frac{H^{\varphi}_\ell + H^{\varphi}_{\ell+1}}{2}
\end{equation}
\begin{equation}
\label{eq:bc2b-tm} 
H^{\varphi}_{\ell+1} - H^{\varphi}_\ell = \sigma_{e(\ell)} \frac{E^{z}_\ell + E^{z}_{\ell+1}}{2}
\end{equation}

\subsection{Effective Medium Theory}
\label{app2:EMT}
For the CPA configuration of Fig. 1(b), i.e. the coated multi-layer cylinder embedded in the effective medium, the scattering coefficients can be derived through the equation for the total transfer matrix, which reads as

\begin{equation}
\mathbb{M}_{(N+1),\nu}
\begin{pmatrix}
b_\nu \\[6pt]
0
\end{pmatrix} = 
\begin{pmatrix}
1 \\[6pt]
a_\nu^{(N+2)}
\end{pmatrix}
\end{equation}
$\mathbb{M}_{(N+1),\nu}$ is the total transfer matrix of the layered infinitely-long cylinder with $N+1$ layers.

Therefore, the coefficient $b_\nu$ and the scattering coefficient $a_\nu^{(N+2)}$ will be
\begin{equation}
b_\nu = \frac{1}{\mathbb{M}_{(N+1),\nu}^{(11)}}
\end{equation}
\begin{equation}
 a_\nu^{(N+2)} = \mathbb{M}_{(N+1),\nu}^{(21)}b_\nu = \frac{\mathbb{M}_{(N+1),\nu}^{(21)}}{\mathbb{M}_{(N+1),\nu}^{(11)}}
\end{equation}
As stated in the main text, in order for an incoming wave to see a truly homogeneous medium, the scattering coefficient $a_\nu^{(N+2)}$  has to vanish (generalized CPA equation):
\begin{equation}
 a_\nu^{(N+2)} =  \mathbb{M}_{(N+1),\nu}^{(21)}b_\nu = \frac{\mathbb{M}_{(N+1),\nu}^{(21)}}{\mathbb{M}_{(N+1),\nu}^{(11)}} = 0
\end{equation}
which leads to: 
\begin{equation}
\label{eq:Generalized-CPA}
\mathbb{M}_{(N+1), \nu}^{(21)}= 0.
\end{equation}

The above $\mathbb{M}_{(N+1), \nu}$ transfer matrix can be written as 
\begin{equation}
    \mathbb{M}_{(N+1),\nu} = \prod_{\ell=N+1}^{1}  \mathbb{T}_{
    \ell\nu} = \mathbb{T}_{(N+1),\nu}\prod_{\ell=N}^{1}  \mathbb{T}_{\ell\nu}
\end{equation}
For simplicity we will denote $\mathbb{B}_{(N),\nu}=\prod_{\ell=N}^{1} \mathbb{T}_{\ell\nu}$. 
This matrix contains information only about the original system. 
Using index notation for matrix multiplication $C^{ik} = \sum_{j} A^{ij} B^{jk}$ the generalized-CPA equation will be 
\begin{equation}
    \mathbb{M}_{(N+1),\nu}^{(21)} = 0 =  \mathbb{T}_{(N+1),\nu}^{(21)} \mathbb{B}_{(N),\nu}^{(11)} + \mathbb{T}_{(N+1),\nu}^{(22)} \mathbb{B}_{(N),\nu}^{(21)} 
\end{equation}
or
\begin{equation}
    -\frac{\mathbb{T}_{(N+1),\nu}^{(21)}}{\mathbb{T}_{(N+1),\nu}^{(22)}} = \frac{\mathbb{B}_{(N),\nu}^{(21)}}{\mathbb{B}_{(N),\nu}^{(11)}}
\end{equation}
For the term appearing in the left-hand-side we have  $a_\nu(R_{N+1};\eff,h)=-\mathbb{T}_{(N+1),\nu}^{21}/\mathbb{T}_{(N+1),\nu}^{22}$, which is equal to the scattering coefficient of a single cylinder with electric permittivity $\varepsilon_\eff$, magnetic permeability $\mu_\eff$ and radius $R_{N+1}$ embedded in a host with electric permittivity $\varepsilon_h$ and magnetic permeability $\mu_\eff$. The right-hand-side is equal to $a^{(N)}_\nu=\mathbb{B}^{21}_{(N),\nu}/\mathbb{B}^{11}_{(N),\nu}$ (scattering coefficient of N-layer cylinder in the host).  

\subsection{Extinction efficiencies for $N=2$ and $\sigma_e=0$ or $\sigma_m=0$}
\label{app:Zero-conductivity}
Here we present the case of a double-layered cylinder coated with a metasurface of either $\sigma_e=0$ or $\sigma_m=0$. We keep $R_1=100$~nm, $f_e=21$~THz and $f_m=18$~THz.
The extinction efficiencies are shown in Fig.~\ref{fig:Qext-2L-ZeroCond}.
Comparing these results with Fig.~\ref{fig:Huygens-Qext}(c)-(d) of the main text one can see several differences.
Starting from the case with $\sigma_e=0$ [Fig.~\ref{fig:Qext-2L-ZeroCond}(a)], we can see that the electric modes are absent from the extinction spectrum with only the magnetic modes TE$_0$ and TM$_1$ being present.
Interestingly, while the two dipolar TM$_1$ modes observed in Fig.~\ref{fig:Huygens-Qext} remain also here, there is no the second TE$_0$ mode observed for the metasurface with non-zero $\sigma_e$ and $\sigma_m$, verifying that this resonance corresponds to magnetic response in high index dielectrics. 
Analogous is the case of $\sigma_m=0$ where only the electric TM$_0$ and TE$_1$ modes appear in the extinction spectrum. 

\end{document}